\documentstyle[12pt]{article}
\begin{document}

\def\lsim{\mathrel{\rlap{\lower3pt\hbox{\hskip0pt$\sim$}}
    \raise1pt\hbox{$<$}}}         
\def\gsim{\mathrel{\rlap{\lower4pt\hbox{\hskip1pt$\sim$}}
    \raise1pt\hbox{$>$}}}         

\begin{titlepage}
\renewcommand{\thefootnote}{\fnsymbol{footnote}}

\begin{flushright}
TPI-MINN-95/9-T\\
UMN-TH-1339-95\\
UND-HEP-95-BIG\hspace*{.01em}05\\
hep-ph/9505397\\
\end{flushright}
\vspace{.3cm}
\begin{center} \LARGE
{\bf b $\rightarrow$ s + $\gamma$: A QCD Consistent Analysis of the
Photon Energy Distribution}
\end{center}
\vspace*{.3cm}
\begin{center}
{\Large
R. David Dikeman$^a$, M. Shifman$^a$ and N.G. Uraltsev$^{a,b,c}$
\\
\vspace*{.4cm}
{\normalsize
{\it $^a$ Theoretical Physics Institute, Univ. of Minnesota,
Minneapolis, MN
55455}}\\
{\normalsize
{\it $^b$ Physics Dept., Univ. of Notre Dame du Lac, Notre Dame, IN
46556}}\\
{\normalsize
{\it $^c$ Petersburg Nuclear Physics Institute, Gatchina, St.
Petersburg
188350}}\\
\vspace*{.2cm}
e-mail addresses:\\
{\it \normalsize dike0003@gold.tc.umn.edu},
{\it \normalsize shifman@vx.cis.umn.edu}\\
\vspace*{.95cm}}
{\Large{\bf Abstract}}
\end{center}
\vspace*{.3cm}
The photon energy distribution in the inclusive $b\rightarrow
s+\gamma$
transitions is a combination of two components: the first
component, soft
physics, is determined by the so called primordial distribution
function,
while the second component, perturbative physics, is governed by
the hard
gluon emission. A simple {\em ansatz} is suggested for the primordial
distribution function which obeys the QCD constraints known so far.
We then
discuss in detail how the hard gluon emission affects the energy
distribution.
An extension of the Sudakov approximation is worked out
incorporating the
Brodsky-Lepage-Mackenzie prescription and its generalizations.
We explicitly calculate the marriage of nonperturbative with
perturbative effects in the way required by OPE, introducing separation
scale $\mu$. A few
parameters, such as $m_b$ and $\mu_\pi^2$ affect the shape of the
distribution and, thus, can be
determined by matching to the experimental data. The data, still
scarce,
while not giving precise values for these parameters,  yield
consistency with
theory: the  current values of the above parameters lie within
experimental
uncertainty. On the theoretical side we outline a method allowing one
to go beyond the practical version of OPE.

\end{titlepage}

\newpage
\section{Introduction}

The inclusive $b\rightarrow s+\gamma$ transition is one of a few
low-energy
processes which is potentially sensitive to new physics. Although the
recent
observation of this transition \cite{CLEO}  has so far detected no
deviations
from the Standard Model expectations, it is not ruled out that
contributions
to the decay parameters
going beyond the Standard Model will be discovered in the future
\cite{hewett}.
Therefore,
untangling the effects of the electroweak interactions from those
of the
strong interactions in this problem becomes an important and urgent
task.

The purpose of this work is to develop a theory which is as close to
fundamental QCD as possible. Some model dependence is,
unfortunately,
unavoidable in the photon energy distribution at the present level of
the strong interaction theory. Our approach in this aspect is to
provide
guidelines that can readily
accomodate
future refinements, if necessary, rather
than giving
a final prescription. In other words, we try to set a
general
framework for the theoretical analysis of $b\rightarrow s\gamma$
and similar transitions.

The basis of our approach is the heavy quark expansion, a version of
the
Operator Product Expansion (OPE)\cite{Wilson}. This method has
experienced a
dramatic development over the last decade (see Ref. \cite{Shifman1}
for a
review and a list of references). The total inclusive widths of heavy
flavor
hadrons are most directly amenable to calculations within the
$1/m_Q$ expansion \cite{early,chay}.
In particular, it was shown that the nonperturbative corrections to
the total
widths start from terms $\sim 1/m_Q^2$ \cite{BUV}. For the
radiative decay
$b \rightarrow   s+\gamma$ they were first calculated in
Ref.~\cite{BBSUV}
and turn out to be rather small. The absence of the $1/m_Q$
corrections does not apply, however, to the shape of the photon
spectrum which depends in a
crucial way on  details of the bound state, {\em viz.}, on the
nonrelativistic ``Fermi motion" of the heavy quark inside the
decaying heavy hadron \cite{prl}.
This nonrelativistic `jiggling' leads to  essential nonperturbative
effects
in the kinematical region near the endpoint, of the relative width
$1/m_Q$,  in accord with the
naive expectations following from the nonrelativistic picture.
However, the
way this ``Fermi motion" manifests itself in QCD is essentially
different
\cite{prl} from the description appearing in the nonrelativistic
picture. A
general formalism
appropriate for treatment of the endpoint domain of the spectrum in
QCD was suggested in Refs.~\cite{JR,matthias,motion}.
This formalism
was applied to the analysis of nonperturbative effects shaping the
endpoint photon spectrum in the transition
$b \rightarrow  s \gamma$ in Refs.~\cite{motion,wise,neubert}.
The results can be summarised as follows: the spectrum is
expressible in terms of
a universal (one-dimensional)
distribution function whose moments are related to local
heavy quark operators of increasing dimensions. This distribution
function replaces the Fermi motion wavefunction of
the
nonrelativistic quark model. In contrast with the latter, the
properties of the
distribution function crucially depend on whether the final state
quark in
the process at hand is light or heavy \cite{motion}. If the $s$ quark
mass is
neglected the distribution function we deal with in $b\rightarrow
s\gamma$
is the light-cone one.  The physical photon
spectrum is obtained as a convolution of the primordial (soft)
distribution function with the perturbative spectral function
corresponding to the hard gluon emission.

Attempts to account for the  motion of the heavy
quark
inside heavy hadrons at a purely phenomenological level were made
previously (e.g. \cite{Ali,ACM}). Models of this type \cite{F1} treat
the heavy
hadron as a bound state of the heavy quark $Q$ plus a spectator,
with a certain momentum distribution.
Although at
first
sight this approach, heavily relying on the nonrelativistic picture,
contradicts QCD, a closer inspection shows \cite{roman} (see also
\cite{Randall}) that  the models of Refs.~\cite{Ali,ACM} are
compatible
with QCD through the leading order in $1/m_Q$
for the decays into light quarks.
The corresponding {\em ansatz}  reduces to a  specific choice of the
primordial distribution function. In
particular, the Gaussian distribution over the Fermi momentum in
conjunction
with the energy-momentum conservation relation of the quasifree
type  (routinely
postulated in phenomenological models) implies  the so-called
``Roman''
function for the primordial distribution \cite{roman}
\begin{equation}
\Phi({x})\;=\; \frac{1}{N}\:
\theta(1-{x} )\,
{\rm e}^{-\left(\alpha (1-{x})-\frac{\beta}{1-
{x}}\right)^2}\;\; ;
\label{roman}
\end{equation}
the dimensioneless parameters $\alpha$ and $\beta$ are certain
functions
of
$m_{\rm sp}^2/p_F^2$ where $m_{\rm sp}$ and $p_F$ are
parameters
of the
AC$^2$M$ ^2$ model \cite{ACM}.

The models of Refs. \cite{Ali,ACM}, being compatible with QCD in
principle,
are still unsatisfactory for a number of reasons. First, the
Roman distribution
function, taken at its face value, may have  problems
accomodating  the values of the
basic QCD
parameters $\overline{\Lambda}$ and $\mu_\pi^2$, as they emerge
from the recent estimates (see below).  Second, these models {\em
per se} are responsible only for a soft smearing of the photon
spectrum. Effects due to the hard gluon emission in the decay process
that smear the spectrum further have to be incorporated
additionally.
The proper combination of  the soft and hard effects in the spectrum
crucially depends on the way one  introduces the normalization point
$\mu$, an aspect which is usually totally ignored in the
corresponding discussions. Finally,  some technical details in the
application of the models to
analysing  experimental data were not completely
consistent
with the heavy quark expansion (see
Ref.~\cite{roman} for details). In this work we address
these issues in turn.

We propose a simple two parameter {\em ansatz} for the
distribution function
consistent with all general
properties expected for the light-cone distribution function
(which is relevant when the heavy quark
decays into  a massless one). Our {\em ansatz}  seems to be more
flexible allowing one to
accommodate
realistic values of the basic QCD parameters. We then dwell on the
main
aspects of inclusion of the hard gluon corrections.
The endpoint behavior of the spectrum is  affected by the
perturbative
radiative corrections which have a double logaritmic enhancement in
the case of
the light final quark (i.e. in the $b\rightarrow s \gamma$ transition).
Although for the actual $b$ quark mass the
logarithm of the
mass is not a particularly large number,  the
presence of  Sudakov's double logarithms enhances the role of the
perturbative gluons.

 An interplay between the perturbative and nonperturbative
smearing of the spectrum is quite peculiar.  A version of the Sudakov
form factor appears that
essentially reduces the absolute height of the spectrum in the
endpoint
region. At the same time, the
{\em shape} of the spectrum  is  formed
mainly by
the primordial distribution function \cite{motion}. Accounting for
the
bremsstrahlung corrections is  crucial for  centering  the
predictions around the realistic values of the hadronic parameters.

Decreasing  the normalization point $\mu$ we change the definition
of which gluons are to be treated as soft and which as hard.
Accordingly, the primordial distribution and the hard
perturbative one vary simultaneously -- a part of the primordial
smearing leaks into the perturbative smearing. In principle, the
physical distribution function must be $\mu$ independent, of course.
However, since approximations are made in different ways
in the soft and hard domain, the $\mu$ independence of the
physical distribution function may hold only approximately.
One of our tasks is to check this approximate $\mu$ independence.
To achieve this goal we, obviously, must explicitly introduce the
separation scale $\mu$ in calculating the perturbative distribution
(i.e. the coefficient functions in the Wilson OPE). Thus, we go beyond
the practical version of OPE \cite {NSVZ}.

After the theoretical framework is set we then compare the
theoretical
spectrum with the CLEO data
\cite{CLEO}. The data, however,
are extremely crude at present. Within the existing  uncertainty the
parameters
cannot be precisely attained. Still, though, one can make some
attempts with
the data and show that the current level of analysis is not only
theoretically
consistent, but also consistent with information on the basic QCD
parameters obtained from other sources. It then remains for the data
to improve for the analysis to become  more predictive.

The organization of the paper is as follows. In Sect.~2 we formulate
the
problem and remind the basic elements
of the theoretical description. In Sect.~3 an {\em ansatz} for the
primordial
light-cone distribution function is suggested. Sect.~4
is devoted to perturbative corrections; here we present
both
rather standard expressions and new considerations called upon to
get a more
accurate description in a manner consistent with OPE. In particular,
in Sect. 4.6 a method is outlined allowing one to explicitly
introduce the separation scale $\mu$. Sect.~5
finalizes our
expressions for the total spectrum. In Sect.~6 we briefly discuss
numerical
results and give a  comparison with the first experimenatal data
available now. In Sect.~7 our conclusions  are summarized.

\section{Formulating the Problem}

The $b\rightarrow s\gamma $ transition at the fundamental level is
generated by electroweak penguins \cite{penguins}.  Since in this
work we are
interested only in the strong interaction effects at distances $m_b^{-
1}$ and
larger  we will just assume that the effective Lagrangian governing
this
transition has a local form
\begin{equation}
{\cal L}_{b\rightarrow s\gamma}=\frac{h}{2} F_{\mu\nu}
\bar s (1+\gamma_5) i\sigma_{\mu\nu} b
\label{lag}
\end{equation}
where $h$ is a constant which includes  the electroweak
physics as well as the hard gluon corrections \cite{F2}, both
coming from distances smaller than
$m_b^{-1}$ \cite{HGC}.
We hasten to add that the assumption of locality of the interaction
(\ref{lag})
is not absolutely correct. We will return to this point at the end of
the Section.
We also note that the $s$ quark (current) mass will be consistently
neglected.

The total inclusive decay width is given
by
\begin{equation}
\Gamma(b\rightarrow s\gamma)=
\Gamma_0\left[1+\gamma_1\frac{\alpha_s}{\pi} +
\gamma_2
\left(\frac{\alpha_s}{\pi}\right)^2+...\,+{\cal O}(1/m_b^2)\right]
\label{width}
\end{equation}
where
$$
\Gamma_0 =
\frac{h^2}{4\pi} m_b^3\;\; ,
$$
$\gamma_{1,2,...}$ are numerical coefficients, and
all quantities, including $h$ and $m_b$, are assumed to
be
normalized at the scale $m_b$. The expansion in the square brackets
is well-defined within Wilson's OPE. In particular, all coefficients in
the perturbative
expansion  are infrared finite, and the whole series is
well-behaved. We will not be interested in the  corrections to the
total width which, as can be explicitly seen, are small.

The subject of our study is the photon energy  spectrum which,
unlike the
total width,  is strongly modified by corrections already at the level
${\cal O}(1/m_Q)$. Indeed, in the parton model (valid
in the
academic limit $m_Q\rightarrow\infty$) the adequate description of
the
inclusive transition is given by the two-body decay: a free $b$ quark
decays into a free $s$ quark and a photon. Accordingly, the photon
spectrum is monochromatic
\begin{equation}
\frac{1}{\Gamma_0}\frac{d\Gamma_0}{dE}=\delta\left(E-
\frac{m_b}{2}\right)
\label{monochrom}
\end{equation}
where $E$ is the photon energy.
It is clear that in the actual decay of the $B$ meson the
monochromatic line
at $E=m_b/2$ is smeared, and the photon spectrum stretches
both,
upwards, to the kinematic boundary at $E=M_B/2$, and
downwards,
to zero.

Two distinct effect are responsible for this smearing.
First, the decaying heavy quark is not at rest in the restframe of $B$.
It is
submerged in a soft medium, light cloud, which exchanges energy
and momentum
of order of $\Lambda_{\rm QCD}$ with the
heavy quark at hand. Thus, the soft gluon medium from the cloud
creates a
{\em
primordial}
nonperturbative distribution which is solely responsible, in
particular,
for filling in the window between $m_b/2$ and $M_B/2$
\cite{FONO}. Second,
in the process of decay the quarks can emit and absorb hard gluons,
i.e. those
whose momenta are much larger than  $\Lambda_{\rm QCD}$. For
example,
the $s$
quark can shake off some of its energy and momentum by emitting a
hard gluon which can result in the photon energy lying much below
$m_b/2$. This gluon emission produces a long tail in the photon
spectrum
below $m_b/2$. If $E$ is close to $m_b/2$ a subtle interplay
between
the two mechanisms takes place.   As a matter of fact,
the very
definition of
what can be called `soft gluon' from the cloud (the corresponding
effect will be
described by the primordial distribution function) and what can be
called
`hard gluon' (the effects due to the hard gluon emissions are to be
incorporated
additionally) is a matter of convention. We must introduce a
normalization
point $\mu$, and everything softer than $\mu$ will be referred to
the cloud
while everything harder than $\mu$ will be treated in perturbation
theory.
Correspondingly, all theoretical quantities -- the primordial
distribution
$F(x)$, $\overline{\Lambda}$, and so on --
become $\mu$-dependent.
In principle, the final prediction for the physical quantities (e.g., the
photon spectrum)
should be $\mu$-independent. Since different approximations are
made in the
soft and hard domains, only an approximate $\mu$ independence
will
hold for a physically reasonable choice of $\mu$ (see below).

Technically, the proper way to introduce the primordial distribution
function,
$F(x)$,
with
\begin{equation}
x = \frac {2}{\overline\Lambda}
\left(E - \frac{m_Q}{2}\right)\;\;\;,\;\;\;\overline{\Lambda}=
M_B-m_b\, ,
\label{defx}
\end{equation}
is through consideration of the transition operator. The relevant
operator product expansion for the transition operator runs over
twists, not
dimensions. An exhaustive discussion can be found in
Refs.~\cite{JR} -- \cite{neubert}; here we remind only that
theoretically one predicts
the moments of $F(x)$ as the $B$ meson expectation values of some
local operators, see below. The observed photon spectrum is then
obtained as a convolution of the primordial distribution with the
spectrum appearing at the level of the perturbative gluon emissions,
\begin{equation}
\frac{d\Gamma (E)}{dE}\; = \;\theta(E)\, \int dy F(y)
\frac{d\Gamma^{pert}_b(E-(\overline \Lambda /2) y)}{dE}\;\;.
\label{convolution}
\end{equation}
Integration over $y$ runs from $-\infty$ to 1 (more exactly, the
lower limit of
integration is $y_0=-m_Q/\overline\Lambda$ but we will
consistently ignore
this difference). One should keep in mind that
$d\Gamma^{pert}_b/dE$ is nonvanishing only in the interval
$(0, m_b/2)$. We also note that one can literally use the
perturbative {\em spectrum} in the above equation only as long as
one does
not apply it to the
very low energy part, $E\sim \overline\Lambda$; a more accurate
expression
for this case is discussed in Sect.~5.
Both theoretical components  in Eq.~(\ref{convolution})
will be considered below in detail.

Prior to submerging into the discussion of the nonperturbative and
perturbative distributions, a comment is in order concerning the
locality of the weak vertex in Eq.~(\ref{lag}). For large $m_t$
(we now know that $m_t$ lies in the ballpark of 170 GeV) the
main contribution to the weak vertex  comes from distances
of order $m_t^{-1}$ which are indeed short in the scale $m_b^{-1}$,
so that this part of the vertex is truly local,
and our consideration is applicable in full.
However, some small fraction of the amplitude
(according to estimates of Ref.~\cite{Ali2}, less than $10\%$ of the
total width) comes from the part of the
amplitude associated with the loop momenta $m_c^2 \lsim k^2\lsim
m_b^2$ in the penguin
integral. (The corresponding photon is radiated by the low
momentum virtual
$c$ quark in the penguin graph.)
Numerically this effect is, thus, rather
insignificant; but even this insignificant effect can be properly
treated within the OPE-based
approach as long as one is interested in the endpoint spectrum.

It turns out that the nonlocal part of the decay amplitude
can
modify the absolute decay rate (the endpoint domain  and
the
low $E$ part of the spectrum are modified differently, generally
speaking);
however,  the {\em
shape} of the endpoint spectrum is not affected. This property is the
reflection of a universal nature of the corrections to the endpoint
spectrum which do not depend on the underlying structure of the
decay amplitude, a feature well-known in the theory of  the
Sudakov-like
perturbative corrections. The universality of the nonperturbative
effects has been
discussed in detail in Ref.~\cite{motion}, and the above fact can be
demonstrated by
following a similar line of reasoning. Let us briefly illustrate our
point.

Nonlocality arises due to the gluon emission from `inside' the vertex
(structural gluons). In particular, if the internal quarks (in our case
the
charmed quarks) are relatively light, the amplitude at
the level of
${\cal O}(\alpha_s)$ corrections becomes complex due to real
intermediate
states that become kinematically possible in the decay. And even in
this case
the relation
similar to Eq.~(\ref{convolution}) holds. It rests on two key facts:
first, due to the gauge invariance, the leading twist effects in the
transition amplitude depend only on the {\em total} momentum $k$
of the hadronic system recoiling the photon. In the endpoint region
$k^2\ll m_b^2$
and,
therefore, the momentum $k$ practically does not differ from what
one has at
the parton level (i.e. transition of the free $b$ quark into the free
$s$ quark where $k^2 = 0$ );
in particular, the same light-cone distribution function
$F$ emerges.

The second crucial property is that in
the kinematics at hand the singularities in $k^2$ associated with the
penguin
amplitude itself lie far enough from the physical point
$k^2=m_b(m_b-2E)$ where the transition operator is considered. This
ensures
the following: in considering the transition amplitude in the complex
plane at
small
$k^2$ and performing the operator product expansion (with the aim
of
obtaining the discontinuity of
the transition operator) in the way it was done in
Ref.~\cite{motion}, one picks up only
the contribution of the physical cut corresponding to the decay
channel of
interest, $b\:\rightarrow \:\gamma\,+\,\mbox{strange jet}\,$. The
contributions of other cuts, although yielding a complex decay
amplitude, are
analytical in this kinematical region and have no discontinuity.
In other words, the presence or absence of structural gluons is not
essential
for the spectrum; only the singularities of the perturbative transition
amplitude matter here. A more detailed presentation of these
arguments will be given elsewhere. Here we just note  that one gets
the same physical spectrum  as shown in
Eq.~(\ref{convolution}) even for a nonlocal penguin amplitude -- the
only difference is that ${\rm d}\Gamma_b^{pert}/{\rm d}E$ must be
understood now as the perturbative spectrum generated by the total
penguin
amplitude, including its nonlocal part.

Reiterating,  some corrections due
to nonlocality of the $bs\gamma$ vertex appear in the total
width, in the absolute normalization of the endpoint rate and in the
shape of the lower part of the spectrum. The shape of the endpoint
spectrum
is given by a universal formula.
The lower part of the spectrum or the total width, however, do not
require a particular
care -- they are insensitive to the primordial distribution, nor one
needs to carry out summation of the high-order perturbative
corrections.
Therefore, taking account of the nonlocality of the weak vertex here
can be done in a straightforward manner for both perturbative
\cite{Ali2} and nonperturbative \cite{BBSUV} corrections.

\section{Ansatz for the Primordial Distribution}

In this section we ignore perturbative gluon emissions and discuss
only
the soft heavy quark distribution function which, in this
approximation,  is independent
on the
normalization point. We suggest the following {\em ansatz}
\begin{equation}
F(x)\; = \; \frac{1}{N}\: \theta(1-x)\:
e^{cx}\, (1-x)^\alpha \,\left[1+b(1-x)^k\right]\;\; ,
\;\;\;-\infty < x<1
\label{f}
\end{equation}
where
$$
x = \frac {2}{\overline\Lambda}
\left(E - \frac{m_Q}{2}\right)\;\;\;,\;\;\;\overline{\Lambda}=
M_B-m_b\;\; ,
$$
$E$ is the photon energy; the parameters  $\alpha \, , b \, , c$ are
positive while
$k$ is an integer, $k=1,2,...\;$.
We do not consider the integer $k$ as a fit
parameter; rather several  independent {\em ans\"{a}tze}
corresponding to
different
values of $k$ are introduced. We will focuss here on  the case $k=1$,
and
only briefly comment on  modifications for  $k>1$.

With this choice of the parameters
the function $F(x)$ is positive everywhere in the
physical domain and has exponential
fall-off at large negative $x$. The latter property ensures the
existence of
all moments \cite{F3}. Here and in what follows we will consistently
work in
the leading non-trivial approximation  in
$1/m_b$; all terms which are additionally suppresed by $1/m_b$
are neglected. Then in
the absence of the perturbative  corrections
\begin{equation}
\frac{1}{\Gamma}\frac{d\Gamma}{dE}\;=\;
\frac{2}{\overline{\Lambda}}\,
F\left(\frac{2E-m_b}{\overline{\Lambda}}\right)\;\;.
\label{dG0}
\end{equation}
The primordial distribution function is to be properly normalized. In
particular, for the
first two moments one has
$$
a_0 = \int dx F(x) = 1
$$
\begin{equation}
a_1 = \int dx x F(x) = 0\;\;.
\label{01}
\end{equation}
The first equation determines the normalization factor $N$
whereas the second one
determines the choice of the photon energy reference point to be
equal to
$m_b/2$. The latter can always be satisfied by means of an
appropriate
substitution
$$
(1-x)\;\rightarrow \;\tau (1-x)
$$
which amounts to the change
$$
b\;\rightarrow \;\tau^k b\;\;\;,\;\;\;c\;\rightarrow \;\tau c\;\;.
$$
Therefore, our function in fact has two free parameters: the value of
$\alpha$ and the ratio $\beta\equiv b/c^k$.

In general, our {\em ansatz} for $F(x)$ behaves as an arbitrary
(non-integer) power
of $1-x$ near the kinematical endpoint. This seems to be reasonable
since
purely perturbative bremsstrahlung effects lead, generally speaking,
to a
similar
behavior, as discussed in the next section, and
the
change in the normalization point leads to a change in this power.

The two next moments of the primordial distribution are given
\cite{motion} (see also \cite{Mannel}) by
$$
a_2=\int\; dx \,x^2F(x)=\frac{\mu_{\pi}^2}{3\bar{\Lambda}^2}\, ,
$$
\begin{equation}
a_3=\int\; dx \,x^3F(x)=\frac{1}{6\bar{\Lambda^3}}\frac{1}{2M_B}
\langle B|\sum_q \:g_s^2\:\bar{b}\,t^ab\: \bar q \,t^a\gamma_0q\,
|B\rangle
\label{a23}
\end{equation}
where in the second equation the sum runs over all light quarks and
$t^a$ are
$SU(3)_c$ generators. The matrix element of the kinetic operator is
defined as
$$
\mu_{\pi}^2=\frac{1}{2M_B}\langle B|\bar b (i\vec{D})^2 b
|B\rangle\;\;.
$$

Not much is known at present about even these lowest moments of
$F$.
The best
 theoretical estimate of $\mu_\pi^2$ existing today is derived from
the QCD
sum rules
\cite{p^2} and suggests the value
$$
\mu_{\pi}^2\approx 0.5\,{\rm GeV}^2
$$
(see also \cite{update} for an updated discussion). This estimate is in
good qualitative agreement with the model-independent bounds
\cite{volp,Vcb,optical} and a semiphenomenological estimate of
Ref.~\cite{thirdsr}. For the value of $\overline\Lambda\approx
500\,{\rm MeV}$ \cite{p^2,update,narison} one would
 expect then $a_2\sim 0.7$. The estimate \cite{motion} of the
expectation value of the four-fermion operator in the third moment
leads to $-a_3\sim 0.3$ which, however, is even more uncertain.

One can readily  see that for the function of a reasonable shape it is
not easy to obtain a large value of the second moment. For example,
the
maximal
value of $a_2$ for a symmetric positive function which vanishes for
$|x|>1$ is
$1$,
and if additionally it exhibits only one peak then even $a_2<2/3$.
The actual
distribution function does not have to (and, in fact, cannot)  be
symmetric.
Nevertheless, it is clear that larger values of $a_2\gsim 1$ can be
reached
only at a price of having a relatively pronounced long tail
towards the
negative values of $x$, which, in turn, would imply large negative
values of the
third moment $a_3$ \cite{motion}. In fact,  strict inequalities can
be easily obtained
for any positive function $F$
\begin{equation}
a_2 < \frac{1}{4}+\sqrt{\frac{1}{4}-a_3}\;\;\;,\;\;\;\;\; a_3 <
\frac{1}{4}-
\left(a_2-\frac{1}{2}\right)^2\;\;;
\label{ineq}
\end{equation}
the first inequality is somewhat stronger than the one obtained in
Ref.~\cite{motion}.

To derive these inequalities  one merely observes that for any $t$
the integral over $x$ over the function $(1-x)(x-t)^2 F(x)$ running
from $-\infty$ to $1$  is
positive; on the other hand this integral  is a second order polynomial
in $t$ and,
therefore, its determinant is negative. In a similar manner one can
get inequalities incorporating higher moments.

In view of poor information available about the moments and a
crude nature of experimental data on the photon spectrum we
mainly limit
ourselves to the simplest {\em ansatz} corresponding to $k=1$.
The resulting function for $k=1$
always has a simple one-peak shape.
It turns out that  within this $k=1$ {\em ansatz}
the value
of $a_3$ is strongly correlated with $a_2$, and at fixed $a_2$ a
rather
limited
range of values for $a_3$ can be reached.  For example,
with  $a_2 = 0.5$ we get $ 0.47 < -a_3 < 0.5$. In Fig.~1 we
show the domain of
values of $a_2$ and $a_3$ which are
accessible within the {\em ansatz} of  Eq.~(\ref{f}) with $k=1$ and
$k=3$.
As one can see from the figure,
again at the same point $a_2=0.5$ but for $k=3$
the flexibility of the {\em ansatz} is much better; in fact, at this
point $ 0.29 <-a_3 < 0.55$ if $k=3$.
The limitations imposed by the particular
choice
of the functional form for the primordial distribution  may, under
certain
circumstances, turn out to be important for the phenomenological
extraction of
the
hadronic parameters. In particular, if the  values of
$\overline{\Lambda}$
and $\mu_\pi^2$ are such that they lead to the parameters that lie in
the
vicinity of
the boundary for $a_2$ shown in Fig.~1, one may suspect a
significant
bias
imposed by the
limited flexibility of the model. In this case it will be advantageous to
go beyond this ansatz
to extend the flexibility. This also may be desirable when one has
more
precise experimental data.
The simplest way to modify it is by  considering larger values of $k$,
say
$k=3$.
Then one can easily have less trivially looking primordial
distribution
functions; in particular, it can  have two maxima. Correspondingly,
the
domain of variation of $a_3$ at fixed $a_2$ becomes much wider
which is
illustrated by the dotted line in Fig.~1.
On the other hand, if the
result of the experimental fit happens to fall well inside the allowed
region,
one can assume that the bias caused by the concrete choice of the
distribution
function is minimal.

We pause here to make  the following remark. As it will be discussed
in
detail in Sect.~4, the consistent
treatment of
nonperturbative effects in $1/m_Q$ expansion by means of OPE
requires the
introduction of a  separation scale $\mu$ to ensure that the domain
of
gluon momenta below $\mu$ gets excluded from the
coefficient
functions. In such
a way, in particular, the mass of the heavy quark, $m_Q$ that enters
the
expansion is not the pole mass that cannot be consistently defined at
the level
of nonperturbative effects \cite{anath,pole,bb} but, rather, the
well-defined
running mass \cite{pole};  the parameter
$\overline{\Lambda}$
is  to be understood as  $\overline{\Lambda}(\mu)$. Although this
difference becomes relevant
only when radiative corrections are considered (they are
ignored so far), this fact suggests that the proper numerical value of
$\overline{\Lambda}$ used in the above numbers must be larger by
$\sim$ 100 MeV \cite{pole,volopt} than the estimate
based on the one-loop pole mass routinely used in the applications of
HQET
\cite{HQET}.  Then the
expected
theoretical values of the second moment $a_2$ group around $0.4$
which seems to
be an easy value for our distribution function even at $k=1$.

\section{Perturbative Corrections}

In this Section we forget for awhile about nonperturbative
aspects and discuss the effect of the perturbative QCD
corrections.  The tree-level spectrum is given by the
monochromatic
photon
line at $E=m_b/2\,$, Eq. (\ref{monochrom}). This parton-like formula
is strongly modified by the
gluon
bremsstrahlung.
The  emission of gluons makes the energy distribution continuous in
the whole
kinematical
range $0<E<m_b/2\,$,
still strongly peaking near the endpoint. We will write it in the form
\begin{equation}
\frac{1}{\Gamma}\frac{d\Gamma}{dE}= -\frac{dS}{dE}
\label{spectr}
\end{equation}
where the formfactor \cite{FA}   $S(E)$ is constant outside the
interval
$0 \le E \le m_b/2\,$,
$\;S=1$ at $E<0$ and $S=0$ at $E>m_b/2$.

\subsection{Sudakov's Double Logarithms}

Let us first approach the problem of the gluon emission in the
problem at
hand in
the classic approximation of pure double logs. The inclusive
$b\rightarrow
s\gamma$ transition is an ideal example since the decaying $b$
quark is at
rest while the $s$ quark produced is ultrarelativistic; the
corresponding
drastic rearrangement of the color field manifests itself in producing
two
logarithms for each power of $\alpha_s$ in the loop integrals.
In this way we obtain a reasonable qualitative picture of the
perturbative
spectrum by  summing up all
terms of the form
$\alpha_s^n \ln^{2n}{\left(m_b/(m_b-2E)\right)}$.
The result of the summation is given by the  Sudakov exponent
\cite{PQCD}
\begin{equation}
S_{\rm dl}(E)={\rm e}^{-\frac{2\alpha_s}{3\pi}\ln^2{m_b/(m_b-
2E)}}\;\; .
\label{sudak}
\end{equation}
Correspondingly, in this approximation the photon spectrum reduces
to
\begin{equation}
\frac{1}{\Gamma}\frac{d\Gamma_{\rm
dl}}{dE}\;=\;\frac{8\alpha_s}{3\pi}\,
\frac{\ln{\left(m_b/(m_b-2E)\right)}}{m_b-2E} \,
S_{\rm dl}(E)\;\;.
\label{dlspectr}
\end{equation}
The crucial practical question is what value of $\alpha_s$ is to be
used
in this equation.

In the classical Sudakov approximation the running of $\alpha_s$ is
not seen since it affects only terms beyond the pure double
$\log$ approximation. Moreover, $\mu$ is also not introduced so far.
Below we will try to improve this
approximation by
including, at a certain level, the effects of running.
For the time being  one can use, for example, $\alpha_s$ normalized
at the scale  $m_b$. It is clear that  the double
logarithmic  spectrum in Eq.~(\ref{dlspectr}) is
quite close to the delta function; the width of the peak is of order of
$m_b\exp{(-{\rm const}/\sqrt{\alpha_s})}$.

The above double $\log$ expressions have rather transparent
interpretation
which allows us  to easily improve the precision later on. To this end
let us
recall that the
first order probability of emission of the massless gluon in the
$b\rightarrow s\gamma$ decay
 is
\begin{equation}
\frac{1}{\Gamma}\frac{d^2\Gamma_{g}}{d\omega d\vartheta^2} =
\frac{2\alpha_s}{3\pi\omega(1-\cos{\vartheta})}
\label{firsto}
\end{equation}
where $\omega$ is the  gluon momentum and $\vartheta$ is its
angle relative
to
the momentum of the $s$ quark. In this expression it is assumed that
$\omega\ll
m_b$. The photon energy in the presence of a  gluon in the final state
is given by
\begin{equation}
E = \frac{m_b^2-2\omega (m_b-\omega) (1-\cos{\theta})}
{2m_b-2\omega(1-\cos{\theta})} \simeq
\frac{m_b}{2}-
\frac{k_\perp^2}{4\omega}\;\;  ,
\label{kinem}
\end{equation}
$$
k_\perp \simeq \omega\vartheta \;\; ;
$$
combination of Eq. (\ref{firsto}) and Eq. (\ref{kinem}) reproduces
the first
order continuum part of the photon spectrum
obtained by
expanding  Eq.~(\ref{dlspectr}) in $\alpha_s$.

The full Sudakov  expression for the
formfactor $S$ is obtained in the
following simple way. One starts with
computing the  (first order) probability $w(E)$ for the
gluon to be emitted with such momentum that the photon gets
energy below $E$. This probability is obtained by integrating the
distribution (\ref{firsto}) with
this constraint using
the kinematical relation (\ref{kinem}):
\begin{equation}
w(E)=
\int \:d\omega\,d\vartheta^2\;
\frac{1}{\Gamma}\frac{d^2\Gamma_{g}}{d\omega\,d\vartheta^2} \:
\theta \left(\frac{k_\perp^2}{4\omega}-\left(\frac{m_b}{2}-
E\right)\right) =
\frac{2\alpha_s}{3\pi} \ln^2\frac{m_b}{m_b-2E} \;\; .
\label{w}
\end{equation}
The function $w(E)$ has the meaning  of the probability of  emission
of a sufficiently hard
gluon lowering  the photon energy below $E$. The all-order
summation of
double logs in
the Sudakov formfactor amounts then to  merely exponentiating  this
probability \cite{PQCD},
\begin{equation}
S(E)={\rm e}^{-w(E)}\; ,
\label{exp}
\end{equation}
where $w(E)$ is given in Eq. (\ref{w}). As we will see shortly this
classical result borrowed from QED can be used in QCD only provided
$E$ does
not approach $m_b/2$ too closely \cite{FB}. Otherwise, the gluon
may become too soft, and such gluons, by definition,
must be excluded by the perturbative calculations.

\subsection{The Running $\alpha_s$}

The standard Sudakov line of reasoning does not allow one to collect
subleading terms -- those which have higher powers of $\alpha_s$
for the given number of double logarithms. Needless to say that in
QCD it is practically important to determine an appropriate
scale of
$\alpha_s$ that enters the Sudakov formfactor, which amounts to the
inclusion of
subleading terms. Separation of the subleading terms in the
background of
double logs is a notoriously difficult task. However, one can readily
single out
a class of the subleading terms, namely those which are responsible
for the running of $\alpha_s$, leaving all others aside. A theoretical
justification of this procedure is provided by the fact that $b$,
the first coefficient in the
Gell-Mann-Low function, is a numerically large parameter; those
graphs which reflect the running of $\alpha_s$ do contain powers
of $b$ while all others,
presumably, do not. (In the Coulomb gauge the relevant graphs are
just
the bubble insertions in the gluon line. In arbitrary gauge they can
be traced through their specific dependence on the number of the
quark
flavors). The prescription of keeping only this subset of graphs is the
heart
of the so called BLM approach \cite{BLM}.
In our problem it is natural to combine the BLM prescription
with the Sudakov-type consideration.

The original BLM approach was engineered as a scale-setting
prescription based on the ``$b$-dominance" estimate of the
next-to-leading
correction. We will first see how it works in the problem at hand, and
then
proceed to a generalized scheme which has been used for a long time
in
theoretical discussions of the renormalon divergences of the
perturbative
series, and
was applied recently for numerical improvement of the
perturbative
estimates in QCD \cite{volsmith,BBB,law}.
This generalized approach, as it was used previously, is applicable to
the situations when one
deals with a single gluon line (dressed by all bubbles).   It
amounts
 to inserting the unexpanded expression for the running coupling
constant $\alpha_s (k^2)\sim 4\pi /(b\ln{(k^2/\Lambda^2_{\rm
QCD})} )$
inside the {\em integrand} of a one-loop Feynman graph \cite{F4}
which depends on the gluon momentum
$k$, with the subsequent integration over $k$. In the problem
of the Sudakov formfactor one sums over arbitrary number of
gluon
lines.
However, since the resulting expression is just the exponent of the
one gluon emission (i.e. in the given approximation all emissions are
independent from each other), the application of the generalized BLM
prescription is quite straightforward and legitimate as long as we
stay within the approximation of the exponentiation.

The anatomy of the leading double logarithms presented  above
allows one to  easily carry out the corresponding improvement.
Indeed, for the emission of a soft collinear gluon ($\omega\ll m_b$
and
$\vartheta\ll 1$) the
effective running coupling enters at the scale $k_\perp$, i.e.
$\alpha_s \rightarrow
\alpha_s(k_\perp)$. This fact is well-known in the literature (see,
e.g.
\cite{perp})  and it
can be substantiated as follows. To determine the relevant
momentum
scale involved, let us introduce a fictitious gluon mass $\lambda$ in
the
denominator of the gluon propagator (we will need this technical
trick later on anyway).
We then note that introducing the gluon mass leads to exactly the
same (main) kinematical
impact as the transverse momentum of the gluon; for instance, the
off-shellness of the quark propagator  is given by $k^2\approx
(k_\perp^2+\lambda^2)(m_b/2\omega)$  if the on-shell
gluon is
emitted in the kinematics giving rise to
leading logarithms. In other words, the cutoff over the spacelike
gluon momentum in the rest frame
looks like a cutoff over the transverse momentum if seen in the
infinite momentum frame.

If so, taking  account of the running of $\alpha_s$
one has
$$
w(E)=\frac{4}{3\pi}\;\int
\;\frac{d\omega}{\omega}\,\frac{d\vartheta^2}{\vartheta^2}
\alpha_s(\omega\vartheta)\;
\theta\left(\frac{k_\perp^2}{4\omega}-\left(\frac{m_b}{2}-
E\right)\right)\;\simeq
$$
\begin{equation}
\simeq\; \frac{8}{3\pi}\:\left(
\int_\epsilon^{\sqrt{\epsilon m_b}}\,
\frac{dk_\perp}{k_\perp}\ln{\frac{k_\perp}{\epsilon}}\,\alpha_s(k
_\perp)\,
+\,
\int_{\sqrt{\epsilon m_b}}^{m_b}\,
\frac{dk_\perp}{k_\perp}\ln{\frac{m_b}{k_\perp}}\,
\alpha_s(k_\perp) \right)
\label{run}
\end{equation}
where
$$
\epsilon=m_b-2E\;\; .
$$
The formfactor $S(E)$ is still given by
$\exp{(-w(E))}$. If one neglects the running of $\alpha_s$ then
Eq.~(\ref{run}) obviously reproduces the standard Sudakov
exponent. Eq.~(\ref{run}) has been displayed in the literature in
various forms (see, e.g., \cite{PQCD}).

One can further use the explicit expression (\ref{run}) to estimate
the effective
value of the strong coupling which enters the  Sudakov
formfactor. Expanding
$$
\alpha_s(k_\perp) =\alpha_s(\sigma)-
\frac{b}{2\pi}\alpha_s^2(\sigma)
\ln{\frac{k_\perp}{\sigma}}
$$
one  determines the scale $\sigma$ for which the average of
$\ln({k_\perp}/{\sigma})$ in the logarithimic integral in
Eq.~(\ref{run}) vanishes. It is not dificult to check that
$$
\sigma^2=m_b\epsilon\, .
$$
As a result we arrive at the following expression
\begin{equation}
S(E) = {\rm e}^{-\frac{2\alpha_s(\sqrt{m_b\epsilon})}{3\pi}
\ln^2{m_b/\epsilon}}
\label{slog}
\end{equation}
exactly coinciding with the known next-to-leading formula
\cite{smilga}.

We hasten to add that Eq. (\ref{slog}) is valid only outside the
endpoint domain, i.e. when $\epsilon$ is parametrically larger
than $\overline\Lambda$. (Of course, $\epsilon$ has to be
parametrically smaller than $m_b$ for the logarithmic
approximation to have sense.) The Sudakov formfactor in
Eq.~(\ref{slog}) has  a peculiar
feature. If
$\epsilon \sim \Lambda_{\rm QCD}$ the strong coupling that
enters Eq.~(\ref{slog})
corresponds to the scale $\sigma \sim \sqrt{\Lambda_{\rm
QCD}m_b}\gg
\Lambda_{\rm
QCD}$. It seems, naively, that we are deep inside  the perturbative
domain.
This, however, is
not true.
As soon as $\epsilon$ approaches the scale of $\Lambda_{\rm QCD}$,
the integral
(\ref{w}) defining the exponent of the formfactor includes the
region of the
small gluon momenta (of order of $\Lambda_{\rm QCD}$) where the
perturbative consideration is
inapplicable. This regime calls for the explicit introduction of the
normalization point $\mu$ and must be treated separatly. In the
next section we will do a more careful analysis pertinent to the
endpoint domain.

In view of importance of this fact, let us repeat the statement:
although
the expression for the radiative corrections (\ref{slog}) involves
$\alpha_s(\sqrt{\epsilon m_b})$ and therefore, at first sight, allows
descending down to the heart of the endpoint region, actually
 it is applicable only when $\epsilon \gg \Lambda_{\rm QCD}$
(or $\epsilon \gsim \mu =\mbox{several units}\times\Lambda_{\rm
QCD}$ in the logarithmic approximation) when the
integral in Eq.~(\ref{run})
runs over the perturbative domain. This fact has been
noted already in the first of Refs.~\cite{smilga}.  On the other hand,
as long as $\epsilon
\gsim \Lambda_{\rm QCD}\,$, this asymptotics is correct (such an
extravagant
``step-like applicability'' is the usual consequence of the
logarithmic
approximation). This property of the radiative corrections has been
used in Ref.~\cite{motion}.

Let us stress that the breakdown of the perturbation theory when
$E$ approaches
the endpoint by the distance of the typical hadronic energy scale
signals that
nonperturbative effects must appear in the spectrum when
$\epsilon$ decreases
down to $\sim \Lambda_{\rm QCD}$. In other words, as it usually
occurs the
existence of nonperturbative corrections is indicated by the
perturbation theory itself.

\subsection{The Endpoint of the Sudakov Formfactor}

The description of the domain $\epsilon\sim\Lambda_{\rm QCD}$
(where nonperturbative
effects show up) requires a more accurate treatment of purely
perturbative corrections as well;
in particular,  introduction of the renormalization point $\mu$ is
necessary,
following the Wilson procedure where all momenta below $\mu$ are
excluded from the perturbative part of corrections (for a dedicated
discussion see Refs.~\cite{pole,optical}). The excluded soft
contributions
are referred to the primordial distribution function which, then,  also
explicitly depends on $\mu$; in particular, the local
operators that
determine its moments are normalized at the scale $\mu$
\cite{motion}. It is advantageous to choose the
normalization scale $\mu$ as low as possible  still
ensuring
the perturbative regime $\alpha_s(\mu)/\pi \ll 1$.
Practically, we keep in mind that $\mu$ = several units
$\times\Lambda_{\rm QCD}$.
This necessarily modifies the
standard expressions for the Sudakov formfactor.

Consistent introduction of the separation of large and small
distances to
all orders in perturbation theory is an unsolved  technical
problem. In the approximation of interest, however, the task can be
easily
accomplished in different ways. First,
one may ascribe a fictitious mass in the gluon propagator. This route
is very convenient even in a purely technical aspect \cite{volsmith}
and will be discussed in  Sect.~4.5.
Here for illustrative purposes we will adopt even a simpler method
cutting off all
integrations over the gluon transverse momentum at $k_\perp =
\mu$.  As a matter of fact, such cut off yields the same result for
double
logarithmic integrals. In different contexts similar cut off procedure
was used for the same purpose -- discarding the soft contribution --
e.g. in recent work \cite{DW}.

Then we arrive at
$$
w(E;\mu)\;=\;
\frac{8}{3\pi}\left[\,
\int_\epsilon^{\sqrt{\epsilon m_b}}\,
\frac{dk_\perp}{k_\perp}\ln{\frac{k_\perp}{\epsilon}}\,
\alpha_s(k_\perp)\,
\theta(k_\perp-\mu)
\right.
$$
\begin{equation}
+\,\left.
\int_{\sqrt{\epsilon m_b}}^{m_b}\,
\frac{dk_\perp}{k_\perp}\ln{\frac{m_b}{k_\perp}}\,
\alpha_s(k_\perp)
\theta(k_\perp-\mu) \, \right]\, ,
\label{wmu}
\end{equation}
where $w(E,\mu )$ is the same probability as in Sect. 4.1 with the
gluon
momentum cut off.

For the one-loop $\alpha_s(k_\perp)$ the integrals are readily
calculated  analytically. One has
\begin{equation}
w(E;\mu)=\frac{8}{3\pi} \left(
\frac{2\pi}{b}\right)^2
\left[
\frac{1}{\alpha_s(m_b)} \ln{\frac{\alpha_s(\sqrt{\epsilon
m_b})}{\alpha_s(m_b)}} -
\frac{1}{\alpha_s(\epsilon)}
\ln{\frac{\alpha_s(\epsilon)}{\alpha_s(\sqrt{\epsilon m_b})}}
\right]
\label{22}
\end{equation}
$$
\mbox{ for }\;\;\epsilon\ge \mu \, ;
$$
moreover,
$$
w(E;\mu)=\frac{8}{3\pi}\left(\frac{2\pi}{b}\right)^2
\left[
\frac{1}{\alpha_s(m_b)}
\ln{\frac{\alpha_s(\sqrt{\epsilon m_b})}{\alpha_s(m_b)}} -
\frac{1}{\alpha_s(\mu)} \ln{\frac{\alpha_s(\mu)}
{\alpha_s(\sqrt{\epsilon m_b})}}\; -
\right.
$$
\begin{equation}
\left. - \; \frac{b}{2\pi} \ln{\frac{\mu}{\epsilon} } \left(1 -
\ln{\frac{\alpha_s(\mu)}{\alpha_s(\sqrt{\epsilon
m_b})}}\right)\right]
\label{23}
\end{equation}
$$
\mbox{ for }\;\;\frac{\mu^2}{m_b}\le \epsilon < \mu\; ;
$$
and, finally,
\begin{equation}
w(E;\mu)=\frac{8}{3\pi}\left(\frac{2\pi}{b}\right)^2
\left[
\frac{1}{\alpha_s(m_b)}\ln{\frac{\alpha_s(\mu)}{\alpha_s(m_b)}} -
\frac{b}{2\pi} \ln{\frac{m_b}{\mu}}
\right]
\label{24}
\end{equation}
$$
\mbox{ for }\;\;\epsilon <\frac{\mu^2}{m_b}\;\; .
$$
The photon spectrum is given by
\begin{equation}
\frac{1}{\Gamma}\frac{d\Gamma^{\rm pert}}{dE}\;=\; -\frac{d}{dE}\:
\theta(2m_b-E) \,
{\rm e}^{-w(E;\mu)}\;\;.
\label{25}
\end{equation}

The spectrum implied by Eqs.~(\ref{22}) -- (\ref{25}) has a
peculiar feature:
$w(E)$ does not depend on $\epsilon$ for
$\epsilon < \mu^2/m_b$ and is equal to $w(\frac{m_b}{2};\mu)$
given by
Eq.~(\ref{24}).
Therefore, differentiation in   Eq.~(\ref{25})
leaves us with  the vanishing spectrum in the interval
$$
\frac{m_b}{2}-\frac{\mu^2}{2m_b} < E < \frac{m_b}{2}
$$
plus the delta function corresponding to the two-body decay at the
perturbative
endpoint
$E=m_b/2$; the height of the two-body peak is suppressed by
the
factor
$\exp\{-w(\frac{m_b}{2};\mu)\}$.

This behavior of the spectrum is not
surprising. Indeed, as was mentioned above, introducing a cut off
in $k_\perp$ is kinematically equivalent,  in our logarithmic
approximation,
to ascribing a finite mass $\mu$ to
the gluon. It is quite clear then that
the gluon mass $\mu$ purely kinematically does
not allow to
fill in the
fake window (or {\em miniwindow}) $E > E_0(\mu)=(m_b/2)-
(2m_b)^{-1}
\mu^2$ in the perturbative spectrum. Moreover, simultaneously it
lifts the
absolute suppression of the two-body mode,
characteristic to the classical Sudakov expression.   This miniwindow,
somewhat aesthetically unapplealing feature of the
perturbative spectrum obtained above, is not physically significant,
though,
and could
be actually eliminated, as
discussed
below.

A feature which is, perhaps, more
surprising at first sight, is the $\mu$  dependence of the spectrum
in the region $\mu > \epsilon \gg \mu^2/(2m_b)$
corresponding to
the
invariant mass squared of the recoiling hadronic system much larger
than
$\mu^2$. This is a consequence of the logarithmically enhanced
perturbative
corrections involving  integration over the gluon energies $\sim
{\rm d}\omega /\omega$.
For small $\omega \ll m_b$ giving the main contribution,
the kinematical impact of the gluon mass is enhanced: it is given by
$\mu^2/2\omega$ rather than $\mu^2/m_b$.

Below $(m_b-\mu )/2$ the perturbative photon spectrum does not
depend on the
normalization point, as it should be.

The theoretical prediction for the perturbative
photon spectrum
which we present here  corresponds to consideration of the
leading twist operators only.  Fine details
of the
spectrum, at the energy scale smaller than, or of order $
\Lambda_{\rm
QCD}^2/m_b$, are
shaped by high twist
operators completely ignored in  our analysis. As a
matter of fact,
in order to resolve the fine details of the spectrum at this scale one
needs to analyse an infinite chain of corrections of all possible twists
\cite{Shifman1}, a task obviously going beyond  possibilities of
the
present-day theory.
For reasons explained above we  push
$\mu$ as
low as possible, in practice,  a few units $\times
\Lambda_{\rm
QCD}$. Then, at our level of accuracy we can not
distinguish between the spectrum presented above and the one
where the elastic spike is moved to the left so as to close
miniwindow, or just smeared evenly over  the miniwindow. The
corresponding changes in the moments $\int x^n F(x) dx$
are of order $1/m_b$.
 As long as one keeps the integrals intact to order ${\cal O}(m_b^0)$
one can distort
$w(\epsilon )$ compared to the formula  (\ref{24}) at will. In
particular, one of the ways of eliminating  the elastic spike at
$\epsilon = 0$ is as follows.
At $\epsilon \ll \Lambda_{\rm QCD}$ one
can use arbitrary (growing) $w(\epsilon)$ and not necessarily the
one given
by Eqs.~(\ref{23}), (\ref{24}); assigning $w(\epsilon) \rightarrow
\infty$
at $\epsilon \rightarrow 0$ smears the elastic peak completely.
The prediction for the physical photon spectrum remains intact at
the
level of accuracy we work at in this paper.

\subsection{The Endpoint Domain and the $b$ Quark Mass}

Specifying the theory above, we normalized all
quantities including the weak $bs\gamma$ coupling $h$, $\alpha_s$
and
$m_b$ at some
scale. For the
inclusive
$b$ decays the appropriate scale is $\sim m_b$. Thus, we start
from a high scale mass, say, $m_b(m_b)$. On the other hand,
practical
perturbative calculations are done similar to QED analysis
which is phrased, of course, in terms of the on-shell mass. Literal
application
of our expressions, therefore, implies the use of the pole mass,
$m_b^{\rm
pole}$ in QCD. This
mass
is defined to any finite order in  perturbation theory; in fact, it is a
meaningful
theoretical notion in the theory where a strong coupling regime does
not occur.
Then it is tempting to say that the endpoint of the  spectrum is
shaped by
this pole
mass, which
is higher than $m_b(m_b)$.

However, it is well known \cite{anath,pole,bb} that the true pole
mass cannot
be consistently defined in QCD with the precision we are interested
in,
$\lsim
\Lambda_{\rm QCD}$. This exactly corresponds to the circumstance
explained
above in
detail, that one cannot compute  the perturbative spectrum
 for
$\epsilon \sim \Lambda_{\rm QCD}$ naively, without introducing the
normalization point $\mu$. Of course, the problem does not
show up in
the expressions obtained in the pure double logarithmic
approximation
where
the running of $\alpha_s$ is discarded,  a
``frozen'' value of
$\alpha_s$ is used and $\mu$ is not introduced. In this
approximation there is no  difference
with  QED, and the corresponding value of $m_b$
essentially coincides with the ``one-loop pole mass'' discussed in
Refs.~\cite{pole,optical,upset}.

We  go beyond this approximation, however;
correspondingly, the question arises as to what mass enters
Eqs.~(\ref{spectr}) -- (\ref{exp}) and below.  It is clear that the mass
we
deal with in the endpoint domain in our approach  is
the would be pole mass in the theory
where the gluon exchanges with $|\vec k\,| < \mu$ are excluded, i.e.
the {\em running} mass normalized at $\mu$, $m_b(\mu)$. It is
important
to notice
that $m_b(\mu)$  has
nothing to do with the dimensionally regularized mass
$m_Q(\mu)$ (the latter mass is a well defined theoretical notion
but it is
irrelevant for OPE in the infrared region $\mu\ll m_Q$).
The appearence of the running mass normalized at $\mu$ in all
expressions
for the spectrum in the endpoint domain can be exactly proven in
the BLM
approximation where there are no technical
problems with defining the explicit renormalization prescription. We
have
tacitly assumed the BLM approximation in
the calculations above. For a related discussion see Ref.~\cite{upset}.
The
occurence of  $m_b(\mu)$
justifies the concluding remark  of Sect.~3.

The perturbative spectrum does not extend above
$E=m_b/2$, and the
window between $m_b/2$ and  $M_B/2$ is filled when
nonperturbative corrections are accounted for. However, in the
 OPE-based approach the $b$ quark mass, and,
therefore,
the endpoint of the perturbative spectrum, depends on $\mu$.
Let us dwell on this aspect in more detail.
The dependence of the heavy quark mass on
$\mu$ at $\mu\ll m_Q$ is given by \cite{pole,volopt}
\begin{equation}
\frac{dm_Q(\mu)}{d\mu}\simeq
-c_m\frac{\alpha_s(\mu)}{\pi}\;\;,\;\;\;\;c_m=\frac{4}{3}
\label{30}
\end{equation}
and, therefore, $m_Q$ increases with $\mu$ decreasing. Calculating
perturbative
corrections we must stop at  some $\mu_0\gg \Lambda_{\rm QCD}$.
Then the mass we
``see'' in the endpoint region is $m_b(\mu_0)$ and, in particular, the
{\em perturbative} spectrum
spans up to $E_{\rm max}=m_b(\mu_0)/2$ (we assume that
$\mu_0 \sim\mbox{a few units}\times \Lambda_{\rm QCD}$ and
consistently
neglect the
corrections $\sim \mu_0^2/2m_b$). Assume we can go
perturbatively a little
bit further, down to a smaller value of $\mu$. The  perturbative
physics below
$\mu_0$ will populate the spectrum up to $m_b(\mu)/2$ and,
therefore we will observe the decay events above the original
perturbative ``quark endpoint'' $m_b(\mu_0)/2$.

Formally, the whole physics below $\mu_0$ is now treated in the
$\mu_0/m_b$
expansion and is accounted for by a `soft' distribution function
normalized at the scale $\mu_0$. In particular, it refers now as well
to the region between $\mu$ and $\mu_0$ which is still amenable to
the
perturbative treatment.
Moreover, one can even introduce a perturbative analog of
$\overline\Lambda$,
\begin{equation}
\overline\Lambda^{\rm pert} ={m_b(\mu)}-{m_b(\mu_0)}=
{c_m}
\int_\mu^{\mu_0}\, d\mu'\,\frac{\alpha_s(\mu')}{\pi} \simeq
{c_m}\,
\frac{\alpha_s(\mu)}{\pi}(\mu_0-
\mu)
\;+\;...
\label{31}
\end{equation}
Clearly one cannot, though,  push $\mu$ too low to get an
``estimate'' of the
whole window $\overline\Lambda(\mu_0)=M_B-m_b(\mu_0)$.

This remark shows once again that many  qualitative features of the
nonperturbative dynamics have their seeds already in the
perturbative
expansion itself. Moreover, in QCD it is impossible to distinguish  the
perturbative effects from the  nonperturbative  ones, and one should
rather
speak about separation of  the
short
distance effects  from the long distance ones.

Eqs.~(\ref{22}), (\ref{23}) determine the asymptotic suppression of
the domain  $m_b\gg\epsilon \gg  \Lambda_{\rm QCD}$ when the
heavy
quark mass goes to infinity,
\begin{equation}
S \sim
m_Q^{\frac{8}{3b}\ln{\frac{\alpha_s(\epsilon)}{\alpha_s(m_Q)}} +
{\rm
const}} \rightarrow
m_Q^{ {\rm const}(\epsilon)- \frac{8}{3b}\ln{
\ln{\frac{m_Q}{\Lambda_{\rm QCD}}}}}\;\;.
\label{as1}
\end{equation}
It, thus, formally decreases faster than any finite power of $m_Q$.
More instructive is the behavior of the spectrum  as a function of
$\epsilon$
when $m_Q$ is asymptotically large but fixed.
Eq.~(\ref{23}) shows that the formfactor $S$ at small $\epsilon <
\mu$ behaves like some power of $\epsilon$:
\begin{equation}
S \sim \epsilon^{\frac{16}{3b}\left(\ln{\frac{\alpha_s(\mu)}
{\alpha_s(\sqrt{\mu\cdot m_b})}}-1\right)}
\sim  \epsilon^{{\rm const}(m_b) + \frac{4}{3b}\ln{\alpha_s(\mu)} }
\sim \epsilon^{\epsilon_0(\mu)}
\;\;\,.
\label{as2}
\end{equation}
The critical exponent  $\epsilon_0$ actually {\em depends} on the
normalization point,
rather weakly though, as long as one stays within the domain of
validity of the
perturbative treatment \cite{F5}. The variation of $\epsilon_0$ upon
changing
normalization point $\mu$ is calculable and is given by
\begin{equation}
\mu \frac{d\epsilon_0(\mu)}{d\mu}=-
\frac{4}{3}\frac{\alpha_s(\mu)}{\pi}+
\;{\cal O}(\alpha_s^2(\mu)) +{\cal O}(\alpha_s(\sqrt{m_b\mu}))
\;\;\,.
\label{35}
\end{equation}
Again, one cannot use this expression with too low of a
normalization
point,
$\mu\sim \Lambda_{\rm QCD}$, and, thus,
determine a universal scaling law for the perturbative distribution
function.

The results which were presented in Sect. 4.3 take into account
 all terms of the form
$\left(\alpha_s\ln^2{m_b/\epsilon}\right)^k\cdot
\left(b \alpha_s/\pi\right)^l$. In other words we allow to loosen
the Sudakov logarithms provided that the correction
contains $b$, the first
coefficient in the Gell-Mann-Low function.  Thus, the approximation
we work
with can be called all order BLM-improved
double logarithmic approximation (it will be referred to below as
`natural').
It is  similar to the approximations, often rather sophisticated,
used in applications to problems where the Sudakov-type effects
appear. The
most crucial
difference is the   explicit introduction of the separation scale $\mu$.
Although $b$ is not a true free parameter in  QCD,
its numerical value is rather large, and similar BLM-type
approximations
often
work rather well. What is even more important for us, within the
natural
approximation we are able
to capture  main features of the full
result. Some of them are of a rather subtle
nature  and are often misunderstood in the literature.

The BLM prescription treating $b$ as a formal parameter allows one
to determine
the appropriate scheme which must be used for the strong coupling
--
Eqs.~(\ref{run}),
(\ref{wmu}), (\ref{22}) -- (\ref{24}) and others are valid if
$\alpha_s$
is understood in the $V$ scheme. On the other hand, say, the
precise
kinematical boundaries of integrals in terms of $\epsilon$, $\mu$
and $m_b$ cannot be fixed in this way. For
example, changing everywhere in the integrals determining $w(E)\;$
$\epsilon
\rightarrow 2\epsilon$ or $m_b\rightarrow 2m_b$ does not lead to
the terms of
the form $ \left(\alpha_s
\ln^2{m_b/\epsilon}\right)^k\cdot
\left(b \alpha_s/\pi\right)^l$ summed up within
the natural  approximation; their
consistent computation requires an
honest calculation of the next-to-leading terms and is very involved.

An improvement of the Sudakov-type calculations was considered
in the recent publication \cite{KS}. The results obtained there
provide one with the better accuracy in determining the $m_b$
dependence
of the
spectrum for asymptotically large mass. They can not be directly
applied,
however, to the solution of our problem, since  the separation scale
$\mu$ was
not explicitly introduced . The moments \cite {KS}
we would need to consider require
integrating the anomalous dimensions in the region below the
infrared pole. On
the other hand,  introducing the normalization point $\mu$ the way
we do would  violate the scale
invariance, a crucial component of the method of Ref.~\cite{KS}.

Even leaving this aspect aside, determining the spectrum from the
moments
presented in \cite{KS} is a difficult problem.  In the next section we
propose an alternative approach which can be carried through rather
straightforwardly and is expected to provide sufficiently good
accuracy in
$b\rightarrow s\gamma$ transitions.

\subsection{A Suggestion for Further Improving
$d\Gamma^{pert}/dE$}

Combining the BLM prescription with the classical Sudakov
calculation
is already an improvement, beyond any doubt. However, as we have
just
noted, this natural approximation does not distinguish, say, between
$m_b$ and $2m_b$ in the argument of the logarithms.
The logarithm is far from being asymptotically large in the
$b$ decays, and significant
progress in  experimental data on the
$b\rightarrow s+\gamma$
spectrum will call for  a more refined theoretical
description of the
perturbative part of the spectrum.
One will have to adress two
different
questions, namely further  improvement of the bremmstrahlung
spectrum, and
calculation of corrections to the weak decay vertex itself (which  is
not exactly
local). The latter complication has been briefly
addressed in Sect.~2, and we focus now on further improving the
accuracy of
the bremsstrahlung spectrum.
The consistent summation of all subleading logarithms is clearly an
unrealistic task,
the more so that we want to do it in a way  compliant with the
Wilson
OPE, i.e. discarding the infrared domain in the perturbative
calculation.   Therefore, we are putting forward a suggestion
resulting in what we will call
``the advanced perturbative spectrum" (APS), combining the exact
first-order result, the Sudakov summation of the logarithms and the
extended BLM improvement.

(i) Since $\ln{(m_b/\mu)}$ is not a particularly large parameter we
believe
that it is sufficient to exactly calculate a few first orders in
$\alpha_s$
(in the simplest version -- merely the exact one-gluon expression)
still
continuing the summation of exploding double logarithms.
For $m_b\simeq 4.8\,{\rm GeV}$ the terms ${\cal O}(\alpha_s)$
seem to be more
important numerically than even the next-to-leading logarithm,
${\cal O}((\alpha_s^2/\pi )\ln{(m_b/\mu}))$ (at least when the
latter is not related to the running of the coupling).

(ii)  One should continue to
use the expression for the perturbative spectrum (\ref{spectr}),
\begin{equation}
\frac{d\Gamma^{\rm pert}}{dE}= -\Gamma(\mu)\frac{dS(E;\mu)}{dE}
\label{spectr2}
\end{equation}
with the exponentiated form of the Sudakov formfactor (\ref{exp});
however, the exponent $w(E;\mu)$ is calculated  now as the {\em
exact first order} probability of the
gluon emission within the kinematical constraint that the resulting
photon energy is less than $E$, rather than the double logarithmic
probability as in  Eq.~(\ref{w}).

The parameter $\mu$ must be introduced in such a way as to isolate
and suppress the infrared part.
In the double logarithmic approximation this goal was easily
achieved by discarding the low $k_\perp$ part of the $k$
integration. Now we can do the same job by introducing a fictitious
``gluon mass". The density matrix in the gluon propagator is kept
the same (e.g., in the Feynman gauge it is $\delta_{\alpha\beta}$)
while the Green function $k^{-2} \rightarrow (k^2-\mu^2)^{-1}$.
This ``gluon mass"
does not lead to any bad consequences in the  one-gluon graphs
(exponentiation does not count), which are the
same in the abelian and non-abelian theories; in the abelian
theory the mass term does not violate the
conservation of the (abelian) current. The same trick was proposed
recently in a different context \cite{volsmith,BBZ,BBB}
and we refer the reader to these publications for further comments.

To the first order in $\alpha_s$ the probability of the gluon emission
(leading to the photon energy less than $E$)  is given by
$$
w(E;\mu)\equiv \alpha_s  W(E;\,\mu^2)\;=
$$
\begin{equation}
=\;\frac{1}{\Gamma(\mu)}\,
\frac{1}{64\pi^3m_b}\,\theta(E_0-E)
\int_0^{E}\, dE'\:
\int_{\Delta}^{\Delta(1-\frac{2E'}{m_b})^{-1}}
\: dp_s\;|A(E',p_s,\vartheta)|^2
\label{wexact}
\end{equation}
where
$$
E_0=\frac{m_b}{2}-\frac{\mu^2}{2m_b}\;\;,\;\;\;
\cos{\vartheta}= 1+ \frac{m_b}{E'}\left( \frac{E_0-E'}{p_s}-1 \right)
\;\;,\;\;\;\;
\Delta = E_0 - E' \; .
$$
In the above
equation $E^{'}$ is the  photon energy, $p_s$ is the energy
(momentum)
of the strange quark and $\vartheta$ is the angle between the
momenta of the photon and the quark; $|A(E',p_s,\vartheta)|^2$
denotes
the amplitude squared summed over polarizations of all particles in
the final state~\cite{foot} and color configurations; it depends on
$\mu$ through kinematics. It is directly calculable but the analytic
expression is too lengthy and is not important for our discussion.
We have introduced above $W(E,\mu )$, the probability of having
the photon with energy less than $E$, in the one-gluon
approximation, with the
factor $\alpha_s$ separated out. This quantity will be useful in what
follows.

The total width $\Gamma(\mu)$ is obtained by the exact calculation
of the
decay width within the same ({\em viz.}, first order in $\alpha_s$)
approximation \cite{F6}. The formula for the perturbative spectrum
obtained in this way has
obvious
advantages. On the one hand, it gives the {\em exact}
expression for the one-loop spectrum (with massive gluon)  if
expanded in $\alpha_s$. On the other, it automatically leads to the
summation of the double logarithms near the endpoint region
where the one-loop expressions explode for large  $m_b/\mu$. At
the same time, it respects the exact
relation between the contributions of real and virtual gluons.
This completes the first part of our suggestion.

(iii) The next step of our program that naturally follows from
analysis in
Sect.~4.3 is to BLM-improve this exact one-loop formula by using the
running
strong coupling, thus promoting it to the status of the advanced
perturbative spectrum.
Again, both the exponent $w(E,\mu)$ and the overall normalization
$\Gamma(\mu)$ must be calculated within the same procedure.

Warning: we do not claim that the exponentiation of the exact ${\cal
O}(\alpha_s)$ amplitude (or even of the BLM-improved one)
reproduces the full result in all orders,
as is the case with the standard Sudakov double logarithms.
Moreover, we are
unaware of any adjustable parameters that will control the accuracy
of this
prescription. Nevertheless, arguments can be given that numerically
it must
have a  good accuracy  in the problem at hand. It is important that
the
corrections to exponentiation are proportional to higher powers of
$\alpha_s(\mu)$ (or even evaluated at higher scales), and, therefore,
are to
be calculated completely within perturbation theory.

Calculation of $w(E;\mu)$ in the BLM approximation accounting for
the running
of $\alpha_s$ can be easily done explicitly; it can be written as
the {\em integral} over the fictitious gluon mass of the width
introduced in
Eq.~(\ref{wexact}), i.e. as a simple three-dimensional integral. The
concrete
expression is too cumbersome due to lengthy expression for the
tree-level
decay amplitude itself, and will be given elsewhere. The main
technical
complication is that this BLM-improved calculation must be done
respecting
the separation of low and high momentum scales required by OPE.
The purpose
of the discussion below is to show explicitly how it can be
accomplished in a
straightforward way.

\subsection{Outlining a Method for Introducing $\mu$ beyond the
Leading Logarithmic Approximation}

The extended BLM-improvement of the one-gluon graphs {\em per
se} is a rather simple
technical exercise for Euclidean quantities where one merely uses the
gluon
propagator in the form $\alpha_s(k^2)/k^2$ with the running
$\alpha_s$ inside
the integrand \cite{pole,bb,BBZ,BBB,law}. Moreover, it can be
directly
generalized for
the quantities of the type of decay widths which are formulated
in Minkowski space \cite{volsmith,BBZ,BBB,F8}; the results, however,
cannot
be  used in the all-order resummed form in the both cases
because one
encounters the unphysical infrared pole in the gluon propagator
which has
wrong analytical properties. This problem is absent in the  OPE-based
approach where the contribution of low momentum domain is
excluded
from the perturbative coefficient functions.

Introduction of the hard infrared cutoff is rather trivial in the
extended
BLM approximation for the Euclidean integrals (see
Refs.~\cite{pole,law,upset} for
examples). However, such method is not directly applicable to the
Minkowski integrals. Moreover,  application of OPE to the inclusive
widths
introduces some peculiarity absent in the analysis
of Ref.~\cite{BBB}, and we briefly discuss it here.  Details of the
derivation and the complete discussion of the approach will be given
in a
separate publication \cite{future}.

Thus, our task is to eliminate the soft gluon part in the coefficient
functions.
This task can be achieved in two steps. First, we soften the theory in
the
infrared domain by introducing the gluon mass. In the BLM
approximation the
mass term can be added by hand; otherwise one could have
introduced the
gluon mass through the Higgs mechanism. If this mass is larger than
$\Lambda_{\rm QCD}$ the effective color interaction never becomes
strong.
Then we calculate the coefficient functions in this softened theory.

Even though the theory is now softened in the infrared domain, our
result for
the coefficient function will still contain some (power suppressed)
residual
contributions coming from the infrared domain. We still have to
subtract them,
but now the subtraction is much easier to do than in the unmodified
QCD
because the modified theory is not strongly coupled in the infrared
domain.
The subtraction can be carried out to any given order, and the
procedure
will presumably converge. Implementing this strategy which seems
to be
applicable rather universally (at least in principle), we go beyond the
practical version of OPE \cite{NSVZ}.

A subtle point should be noted immediately. Modifying the theory in
the
infrared domain for calculational purposes we must ensure that it is
{\em not}
modified at large momenta -- otherwise, instead of solving the
technical
problem of building the correct OPE in QCD we will be merely dealing
with a
different theory. In the procedure we suggest this latter requirement
can be
met to any given (fixed) order in the power series.

The advantage of introducing the separation scale $\mu$ in the
above way
over the hard cut off in the Euclidean integrals, with the subsequent
analytic continuation to the Minkowski domain, is rather obvious,
even leaving
aside purely computational aspect.  In doing so we do not violate
the analytical properties of the theory (and, therefore, the
perturbative coefficients themselves) -- the feature which cannot be
achieved if eliminating of the low-energy contribution is carried out
in the ``hard'' way. It is worth emphasizing that the procedure of
softening is  a technical device for OPE-based  calculations. In
particular, the short-distance parts of
the  amplitudes are not required to be unitary -- the unitarity
holds only for the complete physical
amplitude.

How is this general strategy is implemented within the
extended BLM routine? The substitution
$\alpha_s /k^2\rightarrow  \alpha_s  (k^2)/k^2$ for the
renormalon chain in the integrand
leads to the
 Landau pole
at  the (Euclidean) point $\Lambda_{\rm QCD}^2$ where no
singularities are allowed. However, one can merely shift  this pole, by
hand,
to a
Minkowski point $-\mu^2$,
\begin{equation}
\frac{\alpha_s(k^2)}{k^2}\delta_{\alpha\beta}\;\rightarrow \;
\delta_{\alpha\beta}\,\left(
\frac{\alpha_s(k^2)}{k^2}-\frac{4\pi}{b}\frac{1}{k^2-
\Lambda^2_{\rm QCD}}+ \frac{4\pi}{b}\frac{1}{k^2+\mu^2}
\right)\; .
\label{S1}
\end{equation}
Here we use a convention according to which
$k^2$ is Euclidean The one-loop
expression for the running $\alpha_s$ is implied. Here and below all
couplings and $\Lambda_{\rm QCD}$ refer to the $V$ scheme. Of
course, the
large $k^2$ asymptotics is changed, but only at the
level
$k^{-4}$, which is important for power corrections. This drawback
can be  eliminated through any given order in the power
series.

If we want to keep (a finite number) of the power corrections we
must ensure that by modifying $\alpha_s (k^2)/k^2$ in the infrared
domain
we do not spoil the ultraviolet asymptotics, up to the level of the
power terms we are interested in. Then we could choose, for
instance, the infrared regularization as follows:
$$
\frac{\alpha_s(k^2)}{k^2}\delta_{\alpha\beta}\;\rightarrow \;
G^{\rm BLM}_{\alpha\beta}(k^2;\mu)\;=\;
\delta_{\alpha\beta}\,\left(
\frac{\alpha_s(k^2)}{k^2}-\frac{4\pi}{b}\frac{1}{k^2-
\Lambda^2_{\rm QCD}}+ \frac{4\pi}{b}\frac{1}{k^2+\mu^2}   \;+
\right.
$$
\begin{equation}
\left.
+\; \frac{4\pi}{b} \sum_{n=1}^{l}\,
\frac{(\mu^2+\Lambda^2_{\rm QCD})^n}{(k^2+\mu^2)^{(n+1)} }
\right)\;= \; \delta_{\alpha\beta}\,\left[
\frac{\alpha_s(k^2)}{k^2}-\frac{4\pi}{b}\frac{1}{k^2-
\Lambda^2_{\rm QCD}} \left(\frac{\mu^2+\Lambda^2_{\rm
QCD}}{k^2+\mu^2}
\right)^{l+1} \right]
\label{Gblm}
\end{equation}
where $l$ is a finite number corresponding to the number of
Pauli-Villars subtractions.
This function coincides with the unsoftened resummed propagator
$\alpha_s(k^2)/k^2$ at $k^2\gg \mu^2$ up to terms $\sim
\mu^{2l+2}/k^{2l+4}$ and, thus,
allows one to treat properly (within
the accuracy of extended BLM) the condensate corrections through
dimension $2l$ (excluding the dimension of the heavy quark fields
themselves). For instance, to address the effect of the kinetic energy
operator it is {\em a priori} necessary to use $l>1$.
$G^{\rm BLM}(k^2;\mu)$ has no Landau
singularity in the Euclidean domain and is regular and
suppressed at $k^2<\mu^2$; in particular the  effective
coupling vanishes  at $k^2 \rightarrow 0$.

If this infrared regularization method is accepted, the result of
all-order BLM-summation for the generic inclusive width,
$w^{\rm BLM}(\mu)$, (in our case it is $w^{\rm BLM}(E;\mu)$)
can be expressed as an integral over the ``gluon mass'' over the
one-loop value of $w$ obtained with the gluon mass $\lambda$,
$W(\lambda^2)$
with a certain weight function $\phi^{\rm BLM}$:
\begin{equation}
w^{\rm BLM}(\mu)=\int_0^\infty \;\frac{{\rm
d}\lambda^2}{\lambda^2} \;
W(\lambda^2)\, \phi^{\rm BLM}(\lambda^2;\mu)\;\;.
\label{BLMw}
\end{equation}
This expression is similar to the one obtained in Refs.~\cite{BBB};
however,
$\phi^{\rm BLM}$  depends explicitly on the normalization point
$\mu$ and is not a function of only the ``effective gluon mass''
$\lambda$ as was the case in the analysis of Refs.~\cite{BBB}.
Moreover, it has quite different properties. It is given by the
imaginary
part of the gluon propagator in Eq.~(\ref{Gblm}),
$$
\phi^{\rm BLM}(\lambda^2;\mu)= \frac{1}{4}{\rm Im}
\,G^{\rm BLM}_{\alpha\alpha}(\lambda^2;\mu)\;=
$$
\begin{equation}
=\; - \frac{4\pi}{b}
\frac{1}{\ln^2{\frac{\lambda^2}{\Lambda^2_{\rm QCD}} }+\pi^2}
+\frac{4\pi}{b} \delta(\lambda^2-\mu^2)  +
\frac{4\pi}{b}\,\sum_{n=1}^{l}\,
\frac{(\mu^2+\Lambda^2_{\rm QCD})^n}{n!} \delta^{(n)}(\lambda^2-
\mu^2)\; .
\label{phij}
\end{equation}

The first (regular) term in $\phi^{\rm BLM}$, when
substituted in Eq.~(\ref{BLMw}), represents  the
perturbative contribution of the continuum existing for all invariant
masses
$\lambda^2$. The sign of this continuum contribution is negative; this
is a reflection of the asymptotic
freedom. The delta function terms have the right (positive)  sign.
Hence, this continuum contribution is a correction diminishing the
probability.
The $\delta$-like terms in  Eq.~(\ref{phij}) have the meaning of
the
probability of emission of the
gluon with the mass $\mu$ (with small corrections $\sim
\mu^{2n}$
making up for the short distance dependence of this width compared
to the actual massless case). The appearance of such
local terms with higher derivatives is typical
when one applies the OPE procedure to the spectral densities in
Minkowski space \cite{future}.

In principle,  derivation of  Eq.~(\ref{BLMw}) assumes that
$W(\lambda^2)$
is an analytical function of $\lambda^2$ at $\lambda^2\lsim \mu^2$.
This is always
valid  in the consistent application of OPE: a closer singularity
would
imply the presence of other ``soft'' propagators and would mean that
they
have
not been
properly regularized in the infrared by adding the corresponding
operators
in the
OPE. In our particular case of the first order gluon
emission
probability $W(E;\lambda)$, a singularity in the gluon mass
$\lambda$
 occurs
only at
$2m_b-E\simeq\mu^2/m_b$, i.e. only in the region of the
``miniwindow''.
Therefore, this requirement is met.

In the formal limit, $\mu^2\rightarrow 0$,
one essentially reproduces the expressions of
Refs.~\cite{BBB} where in the latter the principal value
prescription is used for treating  the Landau singularity. This
limit can not be taken, however, if one wants to make
OPE-compatible analysis
and to address nonperturbative effects.

The specific form of the infrared softening is not very important.
We have done it by adding $\delta$-functions in $\phi^{\rm BLM}$;
alternatively,
one could have used some smeared functions. The only constraint to
be observed is that the first $l$ moments of $\phi^{\rm BLM}$ must
remain the same. For example, one can consider $\phi^{\rm BLM}$
completely
vanishing in the vicinity of $\lambda^2=0$ which would yield the
propagator analytical at $k^2=0$.
Generically, the gluon Green function in the  Euclidean
becomes then
\begin{equation}
G^{\rm BLM}_{\alpha\beta} (k^2;\mu)=\delta_{\alpha\beta}
\int_0^\infty\;
\frac{d\lambda^2}{\lambda^2} \frac{\phi_j^{\rm
BLM}(\lambda^2;\mu)}{k^2+\lambda^2}\;\;.
\label{gmu}
\end{equation}
As long as the moments are preserved all expressions of this type
are equally
suited for the OPE. Different choices
 it merely correspond to
somewhat
different cutoff procedure, and the matrix elements of the dimension
$n$ operators then differ by  perturbatively calculable
terms $\sim (\alpha_s(\mu)/\pi)^k \mu^n$.

Similar to the approach of Refs.~\cite{BBB}, upon integrating by parts
the integral over the the  gluon mass in  Eq.~(\ref{BLMw})
 can be rewritten in the form resembling
integration over the gluon virtuality with an {\em effective strong
coupling}
$\alpha_s^{\rm eff}$. The quantity that plays the role of the
distribution over the gluon virtuality $\lambda^2$ is
$-{\rm d}W(\lambda^2)/{\rm d}\lambda^2$; however,
$\alpha_s^{\rm eff}$
depends on {\em both} $\mu$ and $\lambda$ and also contains
 local terms:
$$
w^{\rm BLM}(E,\mu)=\int_0^\infty \;{\rm d}\lambda^2 \;
\alpha_s^{\rm eff}(\lambda;\mu)\,
\frac{-{\rm d}W(\lambda^2)}{{\rm d}\lambda^2} \;-
$$
\begin{equation}
+ \frac{4\pi}{b} \left( W(\mu^2)-W(0) \right)  +
\frac{4\pi}{b} \sum_{n=1}^{l}\,
\frac{(-\mu^2-\Lambda^2_{\rm QCD})^n}{n!} \frac{d^n
W(\mu^2)}{(d\mu^2)^n}
\label{BLMw2}
\end{equation}
where
\begin{equation}
\alpha_s^{\rm eff}(\lambda;\mu) = -\int_{\lambda^2}^\infty \,
dk^2 \:
\frac{\phi^{\rm BLM}(k^2;\mu)}{k^2}=
\frac{4}{b}\left[\frac{\pi}{2}- \arctan \left(
\frac{\ln{\frac{\lambda^2}{\Lambda^2_{\rm QCD}} } }{\pi} \right)
\right]\;\;,
\label{eff}
\end{equation}
with $\alpha_s^{\rm eff}(\infty)=0$ and $\alpha_s^{\rm
eff}(0)=4\pi/b$.
One can formally introduce the ``full'' coupling $\tilde{\alpha}_s^{\rm
eff}$
which combines all terms:
$$
\tilde{\alpha}_s^{\rm eff}(\lambda;\mu)\;=\;
\frac{4}{b}\: \left[\frac{\pi}{2}-\arctan\left(
\frac{\ln{\frac{\lambda^2}{\Lambda^2_{\rm QCD}} } }{\pi} \right)
\right]\;
-\frac{4\pi}{b} \theta(\mu^2-\lambda^2) \;-
$$
\begin{equation}
-\;
\frac{4\pi}{b} \sum_{n=1}^{l}
\frac{(\mu^2+\Lambda^2_{\rm QCD})^n}{n!} \delta^{(n-1)}(\lambda^2-
\mu^2)
\label{alphamu}
\end{equation}
which is legitimate for smooth $W(\lambda^2)$; otherwise one
can use a
regular ``smeared'' representation for local terms as was mentioned
above.

For high momenta, when $\lambda^2 \gg \mu^2 \gg
\Lambda^2_{\rm QCD}$, the effective couplings
$\alpha_s^{\rm eff}(\lambda;\mu)$ and
$\tilde{\alpha}_s^{\rm eff}(\lambda;\mu)$ coincide with the
usual strong
coupling $\alpha_s(\lambda)$. However,
$\tilde{\alpha}_s^{\rm eff}$ {\em vanishes} for
small gluon momenta which reflects the
fact of excluding the low momentum (Euclidean) region instead of
assuming a
model
behavior of the gluon propagator in Ref.~\cite{BBB}. One still has to
keep
in mind that all effective couplings introduced above
depend (in the infrared region) on
$\mu$ and, in particular, on the exact way the normalization point is
introduced; on the other hand, the physical results are
scheme-independent,
and therefore the effective couplings cannot have a direct physical
meaning in the low energy domain
whatever technical convenience they grant. Further details will be
available from \cite{future}.

Let us briefly illustrate how relation (\ref{BLMw2}) works in simple
cases.  Consider, for example, the total
inclusive width of a heavy flavor, say,
the process $t\rightarrow b+W\,(+\,\mbox{gluon})$ \cite{volsmith}.
The one loop
correction to the decay
width $\Gamma^{(1)}(\lambda^2)$ decreases fast for $\lambda^2 \gg
m_t^2$ (it does
not vanish for $\lambda > m_t-M_W$ because virtual corrections are
present for
arbitrary finite gluon mass). On the other hand, for
$\lambda^2 \ll (m_t-M_W)^2$ it is practically independent of
$\lambda$.
Therefore with the logarithmic accuracy it can be approximated by
the step-function, $\Gamma^{(1)}(\lambda^2)\simeq \theta(m_t^2-
\lambda^2)
\Gamma^{(1)}(0)$; then one immediately recovers the obvious result
that the
proper momentum scale for the strong coupling in the one loop
correction is
of the order of $m_t$.

To summarize this Section, we believe that the general method
suggested above
can provide one with a very good theoretical accuracy of describing
the
perturbative corrections to spectrum in a number of applications
($b\rightarrow s+\gamma$, semileptonic decays etc.) in the OPE
compliant way.
Let us repeat the necessary steps. First, one calculates the decay rate
of interest to the first order in
$\alpha_s$ with the nonvanishing gluon mass.
(In the case of the photon spectrum
in $b\rightarrow s+\gamma$ one calculates the first order
probability to have photon with energy below $E$).
Then, fixing the normalization point $\mu$ and using
Eqs.~(\ref{BLMw})--(\ref{alphamu}) one sums up all perturbative
corrections
in the BLM approximation to get the ``BLM-improved'' Sudakov
exponent and the
total width which
explicitly depends on the normalization point through the effective
coupling.
Then the decay distribution is given by Eq.~(\ref{spectr2}) which
exponentiates the gluon exchanges. Though looking somewhat
cumbersome, this procedure can be done numerically in a
straightforward
manner.

The very same startegy can be literally applied
to semileptonic decays as well; the only  technical difference is
that here one first needs to fix the invariant mass of
leptons $q^2$ and, at the final stage, integrate over $q^2$. The
contribution of the region of maximal $q^2$ in the
case of $b\rightarrow u$ transitions can be
treated within OPE as described in detail in Ref.~\cite{WA}.

\section{Full $b\rightarrow s+\gamma$ spectrum}

The full observable spectrum is obtained  by convoluting
the perturbative spectrum with the primordial distribution function
responsible for the heavy quark motion due to the soft gluonic
medium. The precise form of the convolution deserves a special
discussion.

An obvious  similarity exists between the classic theory of deep
inelastic
scattering (DIS) \cite{ioffe} and the OPE-based theory of the spectra
in the
inclusive heavy quark decays \cite{JR,matthias,motion}. In  both
cases the expansion
runs over twists, not dimensions, and the predictions are originally
formulated for the moments of certain functions and are given by
matrix elemets of local operators times calculable coefficient
functions (the latter carry all dependence on the large parameter,
$Q^2$ in DIS or $m_Q$ in the heavy flavor decays). The operators
must be
normalized at a scale which stays finite when $Q^2$ or
$m_Q$ tend to infinity. Then the statement of
factorization of the large and short distance contributions in the
moments is
translated into the convolution of the primordial soft function with
the one
describing perturbative evolution due to the hard gluon emissions. In
DIS this is
the essence of the Gribov-Lipatov-Altarelli-Parisi equations
\cite{GLAP}.

However, there are some differences between the two cases
-- in essence, kinematical -- and a straightforward extention of the
parton-model
language and all formulae we are used to in DIS to the theory of the
spectra
in the inclusive heavy quark decays can be unjustified. The main
difference
is as follows. The moments predicted in DIS are the moments of the
structure functions $F(x),\,\,\, 0<x<1$, with respect to the Bjorken
variable $x$.
If we know a large but finite number of moments (i.e. a number not
scaling
with $Q^2$ when
$Q^2\rightarrow \infty$) we know $F(x)$ in
the whole physical domain $0<x<1$ (with a certain accuracy).
Now, in the problem of spectra the appropriate variable $x$ is
defined in Eq.
(\ref{defx}). The knowledge of a finite number of moments with
respect to this $x$ determines the distribution function only in the
window $0<x<1$ and in the nearby endpoint domain $ -1\lsim x< 0$.
One needs to consider the moments
through order $\gsim m_Q/\Lambda_{\rm QCD}$ to address the
whole
spectrum,
including the low $E$ domain,
in the heavy flavor decays. For such high moments the standard
analysis does
not
apply.  Thus, the standard convolution of the primordial and hard
components
(e.g. Eq. (79) in Ref. \cite{motion}) is valid only in the endpoint
domain; what form is valid  for lower values of $E$ is a
question to
be answered.

The reasoning in  Ref.~\cite{motion} was applied to the endpoint
spectrum.
In the present paper, for practical purposes, we need to have  an
expression in
the whole kinematical region $0<E<M_B/2$.
Fortunately, it
is very easy to answer the question what happens outside the
endpoint domain.
We know for a fact, from the OPE analysis,  that at these values of
$E$ the
motion of the
heavy quark inside $B$ affects the physical spectrum only at the
level
of $1/m_b^2$ corrections. Since such corrections are consistently
neglected in the present paper the proper convolution formula
should be
written in such a way that outside the endpoint domain
the physical and perturbative spectra be the same  up to terms
${\cal O}(m_Q^{-2})$.

In fact, the answer can be prompted by a physical picture where one
takes
the $b$ quark at rest and then boosts it to a moving reference frame
to
take account of the primordial motion. This suggests that
\begin{equation}
\frac{1}{\Gamma} \frac{d\Gamma_B}{dE}=\int_{-\infty}^1 \; dx\: F(x)
\cdot \left(1-\frac{\overline
\Lambda}{M_B}x\right)
\frac{d\Gamma_b^{\rm pert}}{dE}\mid_{E'}
\label{41}
\end{equation}
where
$$
E' = E-\frac{\overline
\Lambda}{2}
\frac{2E}{M_B}x\; .
$$
This expression  is obviously equivalent to Eq.~(\ref{convolution}) if
$E$ is
close to $m_b/2$, up to  higher orders in $1/m_b$
distinguishing $M_B$ from $m_b$.
Moreover, both  forms of the convolutions are always equivalent
where the
perturbative spectrum is smooth and can be Taylor expanded.
Say, outside the endpoint domain (but not for $E\sim\Lambda_{\rm
QCD})$ the difference between evaluating the perturbative spectrum
at $E$,
$E'$ or $E-\overline\Lambda x/2$ is noticeable only in the next
order
in $1/m_b$ (and for the convolution, only in $1/m_b^2$) which is
precisely
what we wanted.

The actual difference between Eqs. (\ref{convolution}) and (\ref{41})
can show up only at the points where the perturbative spectrum is
singular.
In principle, it hapens at two points~-- at $E=m_b/2$ (for the
massless $s$
quark) and at $E=0$. At $E=m_b/2$ the formulae
coincide, as was mentioned above. One must, therefore, address only
the point $ E=0$ where the perturbative spectrum contains explicit
$\theta
(E)$ just due to the photon phase space.

Before doing it, let us comment on Eq.~(\ref{41}) from a different
perspective.  This equation
sums up all leading twist contributions  in the expansion of the
transition amplitude,
which shape the endpoint spectrum (for decays
into light quarks) and is formally valid everywhere,
$0<E<M_B/2$. However, outside the endpoint domain the summed
terms are not dominant and
give  corrections of the same order  as some of the higher twist
operators
neglected in the expansion.

Now, consider small $E$. A formal derivation of the full spectrum can
be
carried out using the approach of Ref.~\cite{motion} where the
complete
transition amplitude is considered first in the complex plain. One
then
observes that originally, instead of the perturbative spectrum,
the convolution goes with a certain structure function (the
discontinuity of
the hadronic tensor corresponding to the quark process discribing
the hard
gluon emission). This structure function is directly related, up to
trivial
kinematical factors, to the perturbative spectrum in the physical
domain, but
it is smooth at $E=0$; the vanishing of the perturbative
spectrum  at $E<0$ is ensured by a kinematical factor representing
the
photon phase space which contains $\theta(E)\,$. This
structure function has singularity only near $E=m_b/2$ and, perhaps,
at  some
other distant unphysical points.

Since the structure  function above is smooth at zero, the convolution
with the primordial distribution function in the small $E$ domain
has no effect whatsoever; one just reproduces the perturbative
result. This
is not what is done in Eq.~(\ref{convolution}) which merely ignores
the
phase space factor singularity absent for other values of $E$.
Taking account of this subtlety at small $E$ leads to Eq.~(\ref{41})
which is written in such a way that this effect of the primordial
smearing is just absent at all for small $E$: the smearing goes over
the
interval in $E$ which is itself proportional to $E$; more essentially,
the interval never covers the point $E=0$.
The last factor in Eq. (\ref{convolution}) is introduced {\em ad hoc}
to
ensure the correct normalization (to unity); it is formally of higher
order
in $1/m_b$.

\section{Analysis}

At the present level of experimental data it seems premature to use
APS in the numerical analysis. Thus, we return to the approximation
called ``natural'' in Sect. 4.4 -- exponentiated double logarithms all
order BLM-improved. In other words, in practice we use the
expressions for the perturbative spectrum given by
Eq.~(\ref{25}) with the Sudakov exponent in
Eqs.~(\ref{22})--(\ref{24}).

The form of the convolution exploited to obtain the
theoretical
spectrum of direct photons in $b\rightarrow s+\gamma$ decay is
given in Eq.
(\ref{41}). The value of the strong coupling is taken following from
the one loop value
$\Lambda_{\rm QCD}^{(V)}=300\,\rm MeV$ (in the $V$ scheme)
which nearly
corresponds to $\alpha_s^{V}(2.3\,{\rm GeV}) =
\alpha_s^{\overline{MS}}(1{\rm\, GeV})=0.336$ as suggested by the
most accurate low energy analysis \cite{volmb}.
In view of the poor experimental
data available we limit ourselves to the primordial distribution
functions
with $k=1$ only.

Fig.~2 shows the full spectrum (solid line) obtained for the
theoretically
preferable choice of parameters $\mu_\pi^2=0.5\,{\rm GeV}^2$
\cite{p^2,update}
and $m_b(\mu)=4.67\,{\rm GeV}$ \cite{volmb,volmu} where we use
the normalization
scale $\mu=0.7\,{\rm GeV}$; the dependence of $m_b$ on the
normalization
point is assumed to be given by  Eq.~(\ref{30}) \cite{runmass}.
 The dashed line
shows the
corresponding nonperturbative spectrum which thus is based on
$\overline{\Lambda}(\mu ) =610\,{\rm MeV}$ ($a_2=0.45$
obtained with
$\alpha=0.4$ and $\beta=b/c=6.4$, see
Sect.~3).
We also show purely perturbative
spectrum stemming from Eqs.~(\ref{22})--(\ref{25}) with
$\Lambda_{\rm QCD}^{(V)}=300\,\rm MeV$ and $\mu=0.7\,{\rm
GeV}$;
the area under the $\delta$-function peak constitutes numerically
$0.45$ of the total perturbative width.
It is clearly seen that the account for the perturbative corrections
essentially modifies the resulting spectrum \cite{motion}: first of all,
it
decreases the height of the maximum by a factor of $2.5$, and also
noticeably broadens it.

The exact position of
the normalization scale $\mu=0.7\,{\rm GeV}$ used above was
chosen rather arbitrarily, just to meet best the two usual
requirements -- to
have it as small as possible compared to $m_b$ and, on the other
hand,
still have the perturbative expansion parameter $\alpha_s(\mu)/\pi$
sufficiently small.
Certainly, this still allows a significant variation of this scale. Fig.~3
illustrates the dependence of the full spectrum on the choice of
$\mu$;
however, varying $\mu$ we {\em must}
supplement it by related change in $m_b(\mu)$
in the perturbative spectrum and $\overline{\Lambda}(\mu)$ in the
primordial
distribution function. We used the literal form of Eq.~(\ref{30}) (see
also
Ref.~\cite{upset}) to determine the corresponding masses. In
principle, in
the test of the $\mu$ independence of the physical spectrum
the value of $\mu_\pi^2$ should also have been varied as a function
of
normalization point. We did not do that since numerically the most
important
effect is due to the change of $\overline{\Lambda}(\mu)$.

The solid line in Fig.~3 is obtained with the same, central, choice of
parameters as in Fig.~2; dashed line represents the high normalization
point
$\mu=1\,{\rm GeV}$ and the long-dashed curve shows the case of low
$\mu=0.4\,{\rm GeV}$; the overall normalization is the same for the
three
curves, unity. If $m_b(0.7 \,{\rm GeV})=4.67\,{\rm GeV}$
then the corresponding mass values
for other normalization points are $m_b (1\,{\rm GeV})=4.58\,{\rm
GeV}$ and $m_b (0.4\,{\rm GeV}) =4.83\,{\rm GeV}$.
The value of $\mu_\pi^2=0.5\,{\rm GeV}^2$ was kept fixed
which assumed
rather different values of $a_2$, $0.34$ and $0.83$, respectively.
The normalization point $\mu = 0.4$ GeV
is probably too low to be taken seriously. It is considered only
for illustrative purposes, as an extreme case.

Although both the perturbative spectrum and the primordial
distribution
separately look very different for the three choices of the
normalization
point, the full spectra are very much alike. We stress once more that
it
was crucial to use the proper normalization point dependence of the
heavy
quark mass in order to get this similarity; using a universal,
or ``true'',
mass as sometimes suggested would lead to  essentially different
spectra.
This transparent demonstration of the physical relevance of the
running $b$
quark mass is an exact analog of a similar
example analysed in Ref.~\cite{optical} for the case of the small
velocity (SV) kinematics.

The three curves in Fig.~3 are not exactly identical. A small
difference
emerges for a few reasons. First, it reflects the finite value of the $b$
quark
mass; therefore higher terms in $1/m_b$ neglected in our analysis
give
finite contribution. We also used approximate expression for
the
perturbative spectrum. Second, it is due to the rather limited
flexibility of
the used ansatz for the primordial distribution: in principle, the
functional
form we adopted is not preserved by the renormalization from one
value of $\mu$ to another. This makes it important to use a
physically
justified choice of $\mu$. However, the very fact that the full
spectrum does
appear to depend weakly on $\mu$ indicates that our
approximations work
quite well.

Turning to comparison of theoretical predictions with data
\cite{CLEO}, we
have to admit that the latter are rather crude at the moment to draw
any well
justified definite conclusions. Nevertheless it is fair to say that the
theoretical spectrum for the preferred values of parameters fits well
with what is
seen in experiment. In any case, keeping in mind the fact that the
experimental analysis which lead to the few experimental points we
used,
was rather
involved, their naive fit with any theoretical prediction {\em per se}
is
hardly justified at the moment; the correct conclusion can be made
only
if the comparison to theory is incorporated into the analysis at early
stage,
and, therefore, it
must be left for experiment itself. For this reason in our
present analysis we aim only at tentatative conclusions and rather
try to
illustrate the sensitivity of the spectra to underlying theoretical
parameters; this procedure can give an idea of what kind of precision
one
can expect in determination of these parameters from future
accurate
measurements. In other words, in varying input parametes we, to
some
extent,
take more literally the {\em theoretical} spectrum obtained for the
central
values of parameters as representing the possible experimental
shape, and
compare other spectra with this hypothetical one.

In Fig.~4 we show the full spectrum in its high energy part varying
the $b$
quark mass:  $m_b(0.7\,{\rm GeV})=4.67\,{\rm GeV}$ (solid line),
$\;4.58\,{\rm GeV}$ (dashed  line) and $4.83\,{\rm GeV}$ (long-dashed
line);
these (rather odd at first sight) values are the same as
in Fig.~3 where they merely correspond to
the different choice of the normalization point. We stress that
now the normalization point is fixed at 0.7 GeV. We choose to take
the examples above to separately study  the effects of
independent
varying mass {\em or}
normalization point. The value of $\mu_\pi^2$ is
kept fixed at $0.5\,{\rm GeV}^2$.
Experimental points are also shown, with arbitrary normalization; the
normalization of all three theoretical spectra is the same.

In Fig.~5, on the contrary, we fix $m_b(0.7\,{\rm GeV})=4.67\,{\rm
GeV}$
and try to vary $\mu_\pi^2$
between $0.25\,{\rm GeV}^2$ (dashed  line) and  $0.75\,{\rm GeV}^2$
(long-dashed line).

A look at Figs.~4 and 5 suggests \cite{FC} that at present all
theoretically
allowed
range of parameters is consistent with the CLEO data \cite{CLEO},
and,
apparently, even noticeably wider range of parametes cannot be
excluded.
Still it seems that the most probable values of $m_b$ and
$\mu_\pi^2$ are
likely to lie not far from our current theoretical expectation.
Although, clearly, any more
definite conclusion can emerge only from a dedicated experimental
fit.

In relation to the future data, Fig.~4 shows that the endpoint
spectrum is
rather sensitive to the value of the quark mass, and it seems
probable that
the precise data on the photon spectrum will allow to determine it
here with
the accuracy of about $50\,{\rm MeV}$. On the other hand, according
to
Fig.~5 the dependence of the shape on the expectation value of the
kinetic
operator is not too strong and the possibility to decrease the error in
$\mu_\pi^2$ below $0.15\,{\rm GeV}^2$ looks questionable, if one
takes
into account that the theoretical resolution in the photon energy one
can
pretend on in this approach, is not more than $\sim 50\,{\rm
MeV}$.

\section{Conclusions}

Shortly after it was realized that the parton-model delta function
in the $b\rightarrow s\gamma$ spectrum is substituted
by a universal primordial distrubition in $B\rightarrow X_s
\gamma$ concerns were aired that the hard gluon radiation
of the Sudakov type might
totally wash away the endpoint peak, so that no traces
of the primordial distribution will be visible. The main lesson we
learn
from the present work is that this is not the case. Although the
hard gluon emission definitely smears the primordial function,
when properly treated within OPE this perturbative smearing turns
out to be quite modest. As is seen from the plots discussed in Sect.~5
the peak in the endpoint domain survives perturbative smearing
and is quite conspicuous. We hasten to add that it is not an artefact
of the
finite value of $m_b$ used in the numerical analysis.
The survival of the primordial peak was anticipated in
Ref.~\cite{motion}.

Several reservations must be made as to the accuracy of the various
approximations
we  adopted. Being QCD-compatible, the theory we suggest is
simplified
in many aspects.

First, in the OPE analysis we dealt with the leading twist operators
only.
Higher twist operators will introduce $1/m_b$ corrections, generally
speaking, in all relations
derived above, which may or may not be numerically significant. We
know for
sure, however, that the higher twist operators are absolutely crucial
in a part
of the window, namely in the region which will be called resonance,
$M_B/2 - E \sim \Lambda_{\rm QCD}^2/m_b$. By definition, in this
domain of
the spectrum the hadronic system produced in the radiative $B$
decay is just
one  low-lying strange resonance, like $K^*$.  In the discussion above
we ignored the existence of this region altogether, consistently
repeating
that
the physical spectrum stretches up to $M_B/2$. This was not wrong
in our approximation considering that
the resonance domain is an effect showing up only at the (relative)
resolution  of
$1/m_b^2$. However, since $m_b$ is not academically large in
practical
terms it is impossible to ignore the existence of the resonance region,
at least
for pure kinematical reasons. The mechanism governing the hadronic
transition in the resonance domain may be quite different -- a form
factor type mechanism \cite{chernyak}. Then our prediction for the
physical
spectrum can not  be valid point-by-point on the extreme right of
the
spectrum, when $E$ approaches the kinematic boundary by distance
of order
$\Lambda_{\rm QCD}^2/m_b$. This resonance domain occupies,
roughly,
$\sim 1/5$ of the window.

In general, since only the leading twist operators were considered, in
this
approximation one cannot, in principle, predict  the spectrum with
the energy
resolution of the order of $\Lambda_{\rm QCD}^2/m_b$.
Before confronting
theory and experiment a
smearing over a larger interval of energy must be perfomed. The
smearing is superfluous outside the resonance domain where the
measured
spectrum is smooth by itself. On the other hand, if there is any hope
of
extrapolating the prediction for the spectrum to the resonance
domain, the
prediction will refer to the smeared spectrum, not point-by-point.

With very little information about the high twist contributions, one
can
get an idea about the size of the smearing interval resorting to
consideration of  the impact of the $\rho\,-\,\pi$ or $K^*\,-\,K$ mass
difference
on the kinematics of the decay \cite{motion}. (This
splitting is the most transparent effect
of the next-to-leading twist corrections.) The corresponding
difference
in the
two body decay energy $\delta E$ is given by
\begin{equation}
\delta E  \simeq \frac{M_{K^*}^2-M_K^2}{2m_b}
\simeq \frac{M_\rho^2-M_\pi^2}{2m_b} \simeq 60\,{\rm MeV}\; .
\label{res}
\end{equation}
We see, therefore, that  the size of the resonance domain and the
smearing interval, although formally
scaling like $\Lambda_{\rm QCD}^2/m_b$, is quite significant
numerically.

It may have important practical implications. For example, the
point-to-point
extraction of the distribution function from experimental data in the
region
very close to the kinematic boundary, sometimes proposed for
elimination of
the model dependence of the endpoint determination of $V_{ub}$, is
not likely
to be accurate enough for that reason.

On the other hand, the effect of inclusion of the current quark mass
$m_s$ in
the analysis is minute and can be safely neglected: it scales like
$m_s^2/
2m_b$ (or is linear in $m_s$ in higher twists), and
 is
much
smaller numerically than the effects discussed above.

Addressing  the phenomenological implications of our analysis,
we
report that in spite of the crude nature of the current data for the
photon
spectrum in $B \rightarrow X_s \gamma$, it is still
possible to conclude that the
current theory including both,  perturbative and nonperturbative
effects, is
consistent. Moreover,  the existing theoretical
estimates for $m_b$ \cite{volmb} and $\mu_\pi^2$ \cite{p^2,update}
generate the spectrum which seems
rather close to  the data, rather
than lying near the edge of contradicting them. Bearing in mind a
very crude character of the existing data points we did not attempt
any
$\chi^2$
fit of the data
that would allow us to extract $m_b$, and $\mu_\pi^2$. With data
becoming more accurate such a fit will become a necessity.

We emphasize that our analysis in the future, when more exact data
will appear, will change only in a technical sense.
First, one needs to go beyond the ``natural'' approximation
used in the
present paper and rely on  APS, as it is
described in Sects.~4.5--4.6. The full ${\cal O}(\alpha_s)$
calculation of the effective $b\rightarrow s+ \gamma$ vertex is also
desirable if one wants to consider the whole spectrum and not only
the
endpoint region,
for example, address the total decay width on a parallel footing.
Second, one may need to consider a more general ansatz for the
primordial
distribution
function $F(x)$, say, put $k=3$, which allows a much wider variety of
shapes of the primordial distribution. Then even some experimental
estimates of the third moment may be obtained.

With new precise experimental data one can hope to get a rather
good
independent estimate of the running mass $m_b$ at a relatively
low
normalization point with an accuracy as high as $\sim 50\,\, {\rm
MeV}$.
Also,
a reasonable estimate will be possible for the kinetic matrix element
$\mu_\pi^2$; the accuracy of $ \sim 0.1\,{\rm GeV}^2$ is
conceivable here.
We must note that the  primordial distribution (depending on the
low-energy
parameters) is still rather
significantly smeared by the short-distance perturbative effects.
For this
reason the apparent effect of varying the
low-energy
parameters is softened and their determination would require a
dedicated
analysis. The intervention of short distance corrections is much
less
pronounced in the semileptonic
$b\rightarrow c l\nu$ transitions and, in particular, in the SV
kinematics.
The
latter process,
thus, offers a more accurate and less involoved detemination of
$m_b$ and $\mu_\pi^2$ along the suggestions of
Refs.~\cite{optical,thirdsr}.
The corresponding analysis is under way now \cite{C1}.
Consideration of the semileptonic $b\rightarrow u l\nu$ transitions
is very similar to what is done here for $b\rightarrow s\gamma$.
This
work is also in progress \cite{C2}.
It seems important to make  independent determinations
 in different situations of heavy-to-light and
heavy-to-heavy
transitions where the corresponding distribution functions are
radically
different but are related to each other  by  certain
constraints
\cite{motion}.

{}From the theoretical standpoint, in the present paper we develop the
OPE-based formalism, going explicitly beyond the practical version of
OPE,
using    the inclusive $b \rightarrow s \gamma$ transition as an
example.  For
practical purposes we accepted the
approximation which basically is similar to the double logarithmic
approximation but incorpoprates the BLM-type improvement
accounting for
the
perturbative running of the strong coupling to all orders; we call it
``natural''
approximation. Our consideration seems  instructive in revealing
several   key
elements of
the
OPE-based  approach. One clearly sees here the necessity of
introduction of
the separation scale $\mu$ to discriminate between short-distance
and long-distance effects. The discrimination is rather nontrivial in
the Minkowski kinematics,
even in  calculating the purely perturbative corrections. On the other
hand, it is shown that using the
well defined short distance parameters like $m_b(\mu)$ which do
depend
in the
calculable way on $\mu$, rather than ill defined ``absolute''
parameters of the sort of
``resummed'' pole mass originally suggested in HQET, leads to a
consistent and $\mu$-independent physical spectrum.

We also illustrated that seeds of the basic features of the physical
spectrum -- usually these features
are  attributed purely to  nonperturbative effects --  are
in fact
seen in  the perturbative calculations as long as the
latter
are done in the OPE consistent manner. In particular, we showed that
a  ``window'' which manifests itself as the interval of the
physical
spectrum stretching above the kinematical boundary possible for the
parton level
process, is present already in the proper perturbative treatment. The
consistent application of the ``natural'' expressions also allowed us to
estimate
the asymptotic behavior of the perturbative end point spectrum at
(academically) high mass of the decaying heavy quark, which {\em
is}
$\mu$-dependent as well.

The more accurate description of the spectrum requires going
beyond the
double
logarithmic approximation, even improved by accounting for the
running of the
strong coupling. In particular, the question of the renormalization
point
dependence of the kinetic operator requires specifying explicitly the
normalization scheme; the difference between the schemes, however,
is
beyond the
double logarithmic accuracy in the description of the spectrum.
Numerically such effects constitute about $0.1\,{\rm GeV}^2$ in
$\mu_\pi^2$
\cite{optical}.
We suggest a more advanced and, on the other hand, technically
rather feasible approximation -- the APS. There are good reasons to
believe that APS  must yield sufficient
numerical accuracy: it incorporates
the exact ${\cal O}(\alpha_s)$ result, the summation of the
Sudakov logarithms and, additionally, the running of $\alpha_s$
 within the BLM
hypothesis. The numerical corrections to this approximation,
therefore, are
expected to be rather small  for $b$ decays.

Another new theoretical element of the corresponding analysis is
that
calculation of the perturbative effects is performed
in the way required by OPE,
viz., the perturbative coefficient functions are obtained using
the
explicit infrared cutoff $\mu$. This procedure, more or less
straightforward in the
Euclidean calculations, is rather involved technically in all
Minkowski
processes where spectral densities at nonperturbative level are
addressed. No
consistent approach existed in the literature previously even in the
framework of the BLM
approximation. This problem
must be solved in order to construct APS.
We showed explicitly how it can be done with the example of the
particular regularization scheme which is very convenient in the
BLM-type
calculations; no problems with infrared renormalons and related
ambiguites in
resummation of divergent series appear here as a manifestation of
the
consistent application of Wilson's OPE, and nonperturbative matrix
elements
have definite meaning. This method is applicable in the direct
way to all quantities treated within OPE. In the context of the present
paper
we had to
limit ourselves only to a brief description of the method and
formulating some
results; more details and the discussion of both physical
consequences
and
technical questions will be given elsewhere \cite{future}.
\vspace*{.3cm}

{\bf ACKNOWLEDGMENTS:} \hspace{.4em} The authors are grateful
to G.~Korchemsky, L.~Koyrakh,
A.~Vainshtein and M.~Voloshin for useful discussions.
N.U. gratefully acknowledges discussions with
I.~Bigi, invaluable insights from Yu.~Dokshitzer on bremmstrahlung
effects and exchange of ideas with T.~Mannel about possible ansatz
for
$F(x)$.
This work was supported in part by DOE under the grant
number DE-FG02-94ER40823 and by NSF under the grant number
PHY~92-13313.

\newpage
\vspace*{.5in}

\centerline {\large \bf Figure Captions}
\vspace*{.2in}

\noindent
{\bf Fig.$\,$1} \  The region of the values of $a_2$ and $a_3$
possible with the
{\em ansatz} for $F(x)$ given by
Eq.~(\ref{f}) with $k=1$ (solid line) and $k=3$
(dashed line). We plot the value of $a_3-a_3^{\rm min}|_{k=1}$ along
the $y$
axis, where  $a_3^{\rm min}|_{k=1} \equiv -2a_2^2$.
\vspace*{.2in}

\noindent
{\bf Fig.$\,$2} \ The primordial distribution (dashed line), the
perturbative
spectrum (dotted line) and the full spectrum (solid line) obtained for
$\mu=0.7\,{\rm GeV}$, $\:m_b(\mu)=4.67\,{\rm GeV}$,
$\:\mu_\pi^2=0.5\,{\rm
GeV}^2$ and $\Lambda_{\rm QCD}^{(V)}=300\,\rm MeV$. All curves
shown have the
same area; the height of the elastic peak in the perturbative
spectrum
corresponds to its weight of $0.454$ in the total width. Further
details are
described in Sect.~6.
\vspace*{.2in}

\noindent
{\bf Fig.$\,$3} \ The full end point spectrum obtained at different
values of
the normalization point $\mu=0.7\,{\rm GeV}$ (solid line),
$\:1.0\,{\rm GeV}$ (dashed line) and  $0.4\,{\rm GeV}$ (long-dashed
line);
the running masses $m_b(\mu)$ are used,
$m_b(0.7\,{\rm GeV})=4.67\,{\rm GeV}\,$, $\,m_b(1\,{\rm
GeV})=4.58\,{\rm
GeV}$ and $m_b(0.4\,{\rm GeV})=4.83\,{\rm GeV}\,$. The value of
$\mu_\pi^2=0.5\,{\rm GeV}^2$ is kept fixed.
\vspace*{.2in}

\noindent
{\bf Fig.$\,$4} \ The dependence of the end point spectrum on the
heavy quark
mass; the normalization point is $\mu=0.7\,{\rm GeV}\,$. Solid line is
$m_b=4.67\,{\rm GeV}$, dashed line represents $m_b=4.58\,{\rm
GeV}$ and
long-dashed line corresponds to $m_b=4.83\,{\rm GeV}$. The value
of
$\mu_\pi^2=0.5\,{\rm GeV}^2$ is taken for all curves. CLEO data
\cite{CLEO}
are shown in an aribitrary normalization.
\vspace*{.2in}

\noindent
{\bf Fig.$\,$5} \ The dependence of the end point spectrum on the
expectation
value $\mu_\pi^2$ of the kinetic operator:  solid line is
$\mu_\pi^2=0.5\,{\rm GeV}^2$, dashed line is
$\mu_\pi^2=0.25\,{\rm GeV}^2$
and long-dashed lined is $\mu_\pi^2=0.75\,{\rm GeV}^2$. We assume
$m_b=4.67\,{\rm GeV}$ and  $\mu=0.7\,{\rm GeV}\,$. CLEO data
\cite{CLEO}
are shown in an aribitrary normalization.


\newpage
\begin{thebibliography}{99}

\bibitem{CLEO}
M.S. Alam et al. (CLEO)
{\it First Measurement of the Rate for the Inclusive Radiative
Penguin Decay
$b\rightarrow s\gamma$ }, Cornell preprint CLNS 94/1314; \\
E.H. Thorndike (CLEO), Proc. XXVII International Conference
on High Energy Physics, Glasgow, 1994, Eds. P.J. Bussey and I.G.
Knowles, Institute of Physics Publishing, Bristol 1995, Vol. 2,
page 1327.

\bibitem{hewett}
For a brief review see e.g. J. Hewett, {\it Probing New Physics in B
Penguins},
{\it Preprint} SLAC-PUB-95-6782 [hep-ph/9505247].

\bibitem{Wilson}
K. Wilson, {\it Phys. Rev.} {\bf 179} (1969) 1499; {\it Phys. Rev.}
{\bf D3} (1971) 1818.

\bibitem{Shifman1}
M. Shifman, {\em Theory of Preasymptotic Effects
in Weak Inclusive Decays}, in Proc. of  the Workshop {\it Continuous
Advances
in QCD}, ed. A. Smilga, [World Scientific, Singapore, 1994], page 249
[hep-ph/9405246].

\bibitem{early}
M. Voloshin, M. Shifman, {\it Yad. Fiz.} {\bf 41} (1985) 187
[{\it Sov. Journ.
Nucl. Phys.} {\bf 41} (1985) 120]; {\it ZhETF} {\bf 91} (1986) 1180
[{\it Sov. Phys. -- JETP} {\bf 64} (1986) 698].

\bibitem{chay}
J. Chay, H. Georgi and B. Grinstein, {\it Phys. Lett.} {\bf B247} (1990)
399.

\bibitem{BUV}
I. Bigi, N. Uraltsev and A. Vainshtein, {\it Phys. Lett.} {\bf B293}
(1992) 430; (E) {\bf B297} (1993) 477.

\bibitem{BBSUV}
I. Bigi, B. Blok, M. Shifman, N. Uraltsev and  A. Vainshtein, {\it
The Fermilab Meeting} Proc. of
the 1992 DPF meeting of APS, C. H. Albright {\it et al.}, Eds. World
Scientific, Singapore 1993, vol. 1, page 610; [hep-ph/9212227].

\bibitem{prl}
I. Bigi, M. Shifman, N.G. Uraltsev and  A. Vainshtein, {\it Phys. Rev.
Lett.} {\bf 71} (1993) 496.

\bibitem{JR}
R. Jaffe and L. Randall, {\it Nucl. Phys.} {\bf B412} (1994) 79.

\bibitem{matthias}
M. Neubert, {\it Phys.Rev.} {\bf D49} (1994) 3392.

\bibitem{motion}
I. Bigi, M. Shifman, N.G. Uraltsev and A. Vainshtein,
{\it Int. Journ. Mod. Phys.}, {\bf A9} (1994) 2467.

\bibitem{wise}
A. Falk, E. Jenkins, A. Manohar and M. Wise,
{\it Phys. Rev.}  {\bf D49} (1994) 4553.

\bibitem{neubert}
M. Neubert, {\it Phys. Rev.}  {\bf D49} (1994) 4623.

\bibitem{Ali}
A. Ali, E.~Pietarinen,  {\it Nucl. Phys.} {\bf B 154} (1979) 512.

\bibitem{ACM}
G. Altarelli, N. Cabibbo, G. Corbo, L. Maiani and G. Martinelli, {\it Nucl.
Phys.} {\bf B208} (1982) 365.

\bibitem{F1}
Qualitative discussion of the effects due to the heavy quark motion
inside
heavy hadrons in the problem of the heavy quark fragmentation can
be
traced back to the early days of QCD, see e.g. M. Suzuki, {\it Phys.
Lett.}
{\bf B71} (1977) 139;
V.A. Khoze, Ya. Azimov and L. Frankfurt,
Inv. Talk at the XVIII Int. Conf. on High Energy Physics, Tbilisi, 1976;
J.D. Bjorken, {\it Phys. Rev.} {\bf D17} (1978) 171. A parton-like
approach to
consideration of the distribution functions in the inclusive
semileptonic
$b\rightarrow u$ transitions  was suggested in Ref.~\cite{Paschos}.

\bibitem{Paschos}
A. Bareiss and E. Paschos, {\it Nucl. Phys.}
{\bf B327} (1989) 353; C. Jin, W. Palmer and E. Paschos, {\it Phys.
Lett.}
{\bf B329} (1994) 364.

\bibitem{roman}
I. Bigi, M. Shifman, N.G. Uraltsev and  A. Vainshtein,
{\it Phys. Lett.} {\bf B328} (1994) 431.

\bibitem{Randall}
C. Csaki and L. Randall, {\it Phys. Lett.} {\bf B324} (1994) 451;\\
G. Baillie, {\it Phys. Lett.} {\bf B324} (1994) 446.

\bibitem{NSVZ}
V. Novikov, M. Shifman, A. Vainshtein and V. Zakharov,
{\it Nucl. Phys.} {\bf B249} (1985) 445.

\bibitem{penguins}
The penguin mechanism was introduced in the theory of weak
decays almost exactly 20 years ago, see A. Vainshtein, V. Zakharov
and M.
Shifman
{\it Pis'ma ZhETF}, {\bf 22} (1975) 123 [{\it JETP Lett.} {\bf 22}
(1975) 55]. The purely electromagnetic one-loop penguin in the
theory with two generations relevant to $s\rightarrow d\gamma$
magnetic transition was first discussed in M. Shifman, A. Vainshtein
and V. Zakharov, {\it Phys. Rev.} {\bf D18} (1978) 2583 (operators
$T_{1,2}$ in the nomenclature of this paper). The result has been
generalized to the three-generation model in T. Inami and C.S. Lim,
{\it Prog. Theoret. Phys.} {\bf 65} (1981) 297;
N.G. Deshpande and G. Eilam, {\it Phys. Rev.} {\bf D26} (1982)
2463; B.A. Campbell and P.J. O'Donnel,
{\it Phys. Rev.} {\bf  D25} (1982) 1989; N.G. Deshpande and M.
Nazerimonfared, {\it Nucl. Phys.}
{\bf B123} (1983) 390. From these papers one can extract the
magnetic penguin  for $b\rightarrow s\gamma$ in the Standard
Model.

\bibitem{F2}
Warning: these
genuinely hard gluon corrections should not be mixed with those
which are due to the gluon emissions at distances from $m_b^{-1}$
to $\Lambda_{\rm QCD}^{-1}$. While the former will not be discussed
below the latter are one of the central topics of this paper. We will
reserve
the
term `hard gluon corrections' for the effects of the gluon emissions at
distances from $m_b^{-1}$
to $\mu^{-1}$ where $\mu$ is `the end of the perturbative domain',
$\mu\sim \mbox{several units}\times \Lambda_{\rm QCD}$.

\bibitem{HGC}
The first calculations of the hard gluon corrections were done in
M. Shifman, A. Vainshtein and V. Zakharov, {\it Phys. Rev.} {\bf D18}
(1978) 2583.  Later on the analysis has been repeatedly discussed in
the literature, see e.g. S. Bertolini, F. Borzumati and A. Masiero, {\it
Phys. Rev.
Lett.}
{\bf 59} (1987) 180; N. Deshpande, P. Lo, J. Trampetic,
G. Eilam and P. Singer, {\it  Phys. Rev. Lett.}
{\bf 59} (1987) 183; B. Grinshtein, R. Springer and M. Wise, {\it Phys.
Lett.}
{\bf B202}
(1988) 138; R. Grigjanis, P. O'Donnel, M. Sutherland and H. Navelet,
{\it Phys.
Lett.} {\bf B213}
(1988) 335; (E) {\it Phys. Lett.} {\bf B286}
(1992) 413; G. Cella, G. Curci, G. Ricciardi and A. Vicere, {\it Phys.
Lett.} {\bf
B248} (1990) 181; {\bf B325} (1994) 227;
M. Misiak, {\it Phys. Lett.} {\bf B269} (1991) 161;
M. Ciuchini, E. Franco, G. Martinelli, L. Reina and L. Silvestrini, {\it
Phys. Lett.}
{\bf B316} (1993) 127; {\bf B334} (1994) 137;
M. Misiak, {\it Phys. Lett.} {\bf B321} (1994) 113;
A. Ali and C. Greub, {\it Phys. Lett.} {\bf B259} (1991) 182;
 {\it Z. Phys.} {\bf C60} (1993) 433;
A.J. Buras, M. Jamin, M. Lautenbacher and P. Weisz, {\it Nucl. Phys.}
{\bf
B370} (1992) 69; {\bf  B400} (1993) 37, 75; G. Buchalla and A. Buras,
{\it Nucl. Phys.} {\bf B400} (1993) 225;
M. Ciuchini, E. Franco, L. Reina and L. Silvestrini,  {\it Nucl. Phys.} {\bf
B415}
(1994) 403; A. Buras, M. Misiak, M. Munz and S. Pokorski,
{\it Nucl. Phys.} {\bf B424} (1994) 374.]

\bibitem{FONO}
Let us remind relevant terminology to be used throughout the paper.
The energy interval between the quark kinamtic boundary and the
physical one, $[m_b/2 , M_B/2]$, is called window. Adding to the
window the adjacent domain lying below $m_b/2$, of width
a few units $\times\Lambda_{\rm QCD}$, we get the endpoint
domain.

\bibitem{Ali2}
A.Ali and  C. Greub, {\it Z. Phys.} {\bf C49} (1991) 431;  {\it Phys.
Lett.}
{\bf B287} (1992) 191.


\bibitem{F3}
This is a very important feature. In general,  the primordial
distribution
function is required to fall off faster than any power of $1/x$ at
large
negative $x$. This requirement is due to the fact that the moments of
the
primordial distribution function are related to the matrix elements of
certain well-defined operators. For this reason, {\em ans\"{a}tze}
with a
slower fall off, like, say, that of Ref. \cite{JinP}, have no
interpretation
in QCD as a primordial distribution.

\bibitem{JinP}
C. H. Jin and E. A. Paschos, {\it Preprint} DO-TH 95/07 [hep-
ph/9504375].

\bibitem{Mannel}
T. Mannel, {\it Phys. Rev.} {\bf D50} (1994) 428.

\bibitem{p^2}
P. Ball and V. Braun, {\it Phys. Rev. } {\bf D49} (1994) 2472.

\bibitem{update}
E. Bagan, P. Ball, V. Braun and P. Gosdzinsky,
{\it Phys.~Lett.} {\bf B342} (1995) 362; see also the last of
Refs.~\cite{BBB}.

\bibitem{volp}
M. Voloshin, {\it Preprint} TPI-MINN-94/18-T [unpublished].

\bibitem{Vcb}
M. Shifman, N.G. Uraltsev and  A. Vainshtein,
{\it Phys. Rev.}  {\bf D51} (1995) 2217.

\bibitem{optical}
I. Bigi, M. Shifman, N. Uraltsev and  A. Vainshtein {\it Preprint}
CERN-TH.7250/94 [hep-ph/9405410; to appear in {\it Phys. Rev.}
{\bf D}
(1995)].

\bibitem{thirdsr}
I. Bigi, A. Grozin, M. Shifman, N.G. Uraltsev,
A. Vainshtein, {\it Phys. Lett.}, {\bf B339} (1994) 160.

\bibitem{narison}
S. Narison, {\it Phys. Lett.} {\bf B308} (1993) 365.

\bibitem{anath}
I. Bigi and N.G. Uraltsev, {\it Phys. Lett.} {\bf B 321} (1994) 412.

\bibitem{pole}
I. Bigi, M. Shifman, N.G. Uraltsev and  A. Vainshtein,
{\it Phys. Rev.} {\bf D50} (1994) 2234.

\bibitem{bb}
M. Beneke and V. Braun, {\it Nucl. Phys.} {\bf B426} (1994) 301.

\bibitem{volopt}
M. Voloshin, {\it Phys. Rev.} {\bf D46} (1992) 3062.

\bibitem{HQET}
E. Eichten and B. Hill, {\it Phys. Lett.} {\bf B234} (1990) 511;\\
H. Georgi, {\it Phys. Lett.} {\bf B240} (1990) 447\\
For  recent reviews see e.g. \\
N. Isgur and  M.B. Wise, {\bf in:} {\em $B$
Decays},  S. Stone, Ed.  (World Scientific), Singapore, 1992;\\
H. Georgi,   in  Proceedings of the Theoretical Advanced Study
Institute 1991,  R.K. Ellis, C.T. Hill and J.D. Lykken, Eds.
(World Scientific, Singapore, 1992).

\bibitem{FA}
In fact, $S$ is the square of a
 formfactor; we still will use this somewhat sloppy notation because
the formfactor itself will never appear, so there  can be no confusion.

\bibitem{PQCD}
For a review see e.g. Yu.L.~Dokshitzer, V.A.~Khoze, A.H.~Mueller and
S.I.~Troyan, {\em
Basics of
Perturbative QCD}; Ed. J. Tran Thanh Van,
(Editions Fronti{\`e}res,  Gif-s\"{u}r-Yvette, 1991), and references
therein.

\bibitem{FB}
This fact is demonstrated by Eqs. (\ref{22}) -- (\ref{24}). One should
keep in mind, though, that the latter are more accurate than the
approximation we deal with in the present section.
To make contact with Eqs. (\ref{22}) --
(\ref{24}) one needs to expand in $\alpha_s$ there.
Then Eq. (\ref{22}) coincides with Eq. (\ref{w}) with the additional
constraint $\epsilon >\mu$.

\bibitem{BLM}
S.J. Brodsky, G.P. Lepage and  P.B. Mackenzie, {\it Phys. Rev.} {\bf
D28}
(1983) 228;  G.P. Lepage and  P.B. Mackenzie, {\it Phys. Rev.} {\bf
D48}
(1993) 2250.

\bibitem{volsmith}
B. Smith and  M. Voloshin, {\it Phys. Lett.} {\bf B340} (1994) 176.

\bibitem{BBB}
M. Beneke and V. Braun,  {\it Preprint} UM-TH-94-37 [hep-
th/9411229];
P. Ball, M. Beneke and V. Braun,
{\it Preprint} CERN-TH/95-26 [hep-ph/9502300]; {\it Preprint}
CERN-TH/95-
65 [hep-ph/9503492].

\bibitem{law}
M. Neubert,  {\it Preprint} CERN-TH.7487/94 [hep-ph/9412265].

\bibitem{F4}
Simplified elements of this procedure have been in use
among the Sudakov theory practitioners for years, see, for
example,\\
G.~Altarelli, {\it Phys. Reps.}  {\bf 81} (1982) 1 and further review of
literature therein; Ref.~\cite{PQCD}.

\bibitem{perp}
Yu.L.~Dokshitzer, D.I.~Dyakonov and S.I.~Troyan, {\it Phys.~Reps.}
{\bf 58}
(1980) 270;\\
D.~Amati, A.~Bassetto, M.~Ciafaloni, G.~Marchesini and G.~Veneziano,
{\it Nucl.~Phys.} {\bf B173} (1980) 429.

\bibitem{smilga}
A. Smilga, {\it Nucl. Phys.} {\bf B161} (1979) 449;\\
G. Korchemsky, {\it Phys. Lett.} {\bf B217} (1989) 330;\\
Yu. Dokshitzer, V.A. Khoze and S. Troyan,
 {\it Preprint} LU-TP-92-10, 1992  [unpublished].

\bibitem{DW}
Yu. Dokshitzer and B. Webber,  {\it Preprint} LU-TP 95-8 [hep-
ph/9504219].

\bibitem{upset}
N.G. Uraltsev, {\it Preprint} TPI-MINN-95/5-T [hep-ph/9503404].

\bibitem{F5}
We hope the reader will not be confused by the use of one and the
same
letter, $\epsilon$, for the critical exponent and the photon energy.
The
former is marked by the subscript $0$.

\bibitem{KS}
G. Korchemsky and G. Sterman, {\it Phys. Lett.} {\bf B340} (1994) 96.

\bibitem{BBZ}
M. Beneke, V. Braun and V.I.Zakharov, {\it Phys. Rev. Lett.}
{\bf 73} (1994) 3058.

\bibitem{foot}
We note that if one obtains the gluon density
matrix $\delta_{\alpha\beta}$ (or $\delta_{\alpha\beta}-k_\alpha
k_\beta/k^2$) as a sum over polarizations, for massive gluons one
must sum over three
transverse polarizations rather than two in the massless case.

\bibitem{F6}
In principle, the overall consistency requires the total width to be
calculated with the similar  cutoff $\mu$. This is automatic in the
OPE-based
procedure. The operator expansion for the $b\rightarrow s+\gamma$
total width
is presented  in \cite{BBSUV}; the first non-perturbative terms is
quadratic
in  $\mu/m_b$. Quadratically small corrections are not discused in
the
present paper. Furthermore, in the approximation adopted in the
present paper
the corresponding perturbative matrix elements vanish. Therefore,
one can do
without any normalization point in the calculation of the total width
$\Gamma$  to this accuracy, using for example, the known ${\cal
O}(\alpha_s)$
 expression \cite{Ali2} which is infrared stable.

\bibitem{F8}
A similar consideration for the polarization tensor of light quark
currents was done in M.~Neubert, CERN-TH.7524/94 [hep-ph/9502264].
However, technique used there cannot be
applied to other problems, in particular to heavy flavor transitions.
Let us parenthetically
note that we  disagree with certain aspects of this work.

\bibitem{future}
M. Shifman, N.G. Uraltsev and A. Vainshtein, in preparation.

\bibitem{WA}
I. Bigi, N.G. Uraltsev, {\it Nucl. Phys.} {\bf B423} (1994) 33.

\bibitem{ioffe}
For a review see e.g. B.L. Ioffe, V.A. Khoze and L.N. Lipatov
{\it Hard Proceses}, North Holland, 1984.

\bibitem{GLAP}
V.N.~Gribov and L.N.~Lipatov,
{\it Yad. Fiz.} {\bf 15} (1972) 781; 1218 [{\it Sov. J. Nucl. Phys.} {\bf
15} (1972)] \\
G.~Altarelli, G.~Parisi,  {\it Nucl. Phys.} {\bf B 126} (1977) 298.

\bibitem{volmb}
M. Voloshin,   {\it Preprint} TPI-MINN-95/1-T [hep-ph/9502224,
{\it Int.
J. Mod. Phys.}, to be published].

\bibitem{volmu}
M. Voloshin, in preparation.

\bibitem{runmass}
To make contact with other publications we observe that this
value of the running $b$ quark mass corresponds
to the one-loop pole mass $4.81\,\rm GeV$ if it is extrapolated
from the point $\mu=1\,\rm GeV$.

\bibitem{FC}
Comparing experimental data one also have to keep in mind some
broadening of
the spectrum due to slow movement of decaying $B$ produced in
$\Upsilon(4S)$.

\bibitem{chernyak}
For a review see e.g. V. Chernyak and I. Zhitnitsky,
{\it Phys. Reports}  {\bf 112} (1984) 173.

\bibitem{C1}
L. Koyrakh, N. Uraltsev and A. Vainshtein, to be published.

\bibitem{C2}
H.-L. Yu, privite communication to M.S. and H. Li and H.-L. Yu,
to be published.


\end{thebibliography}
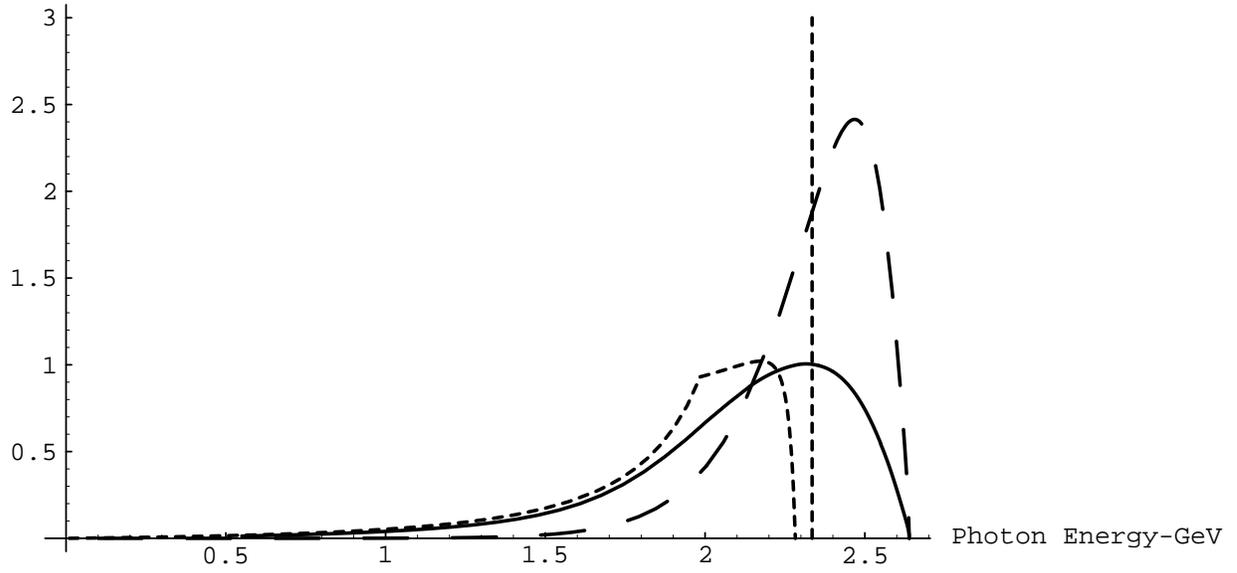
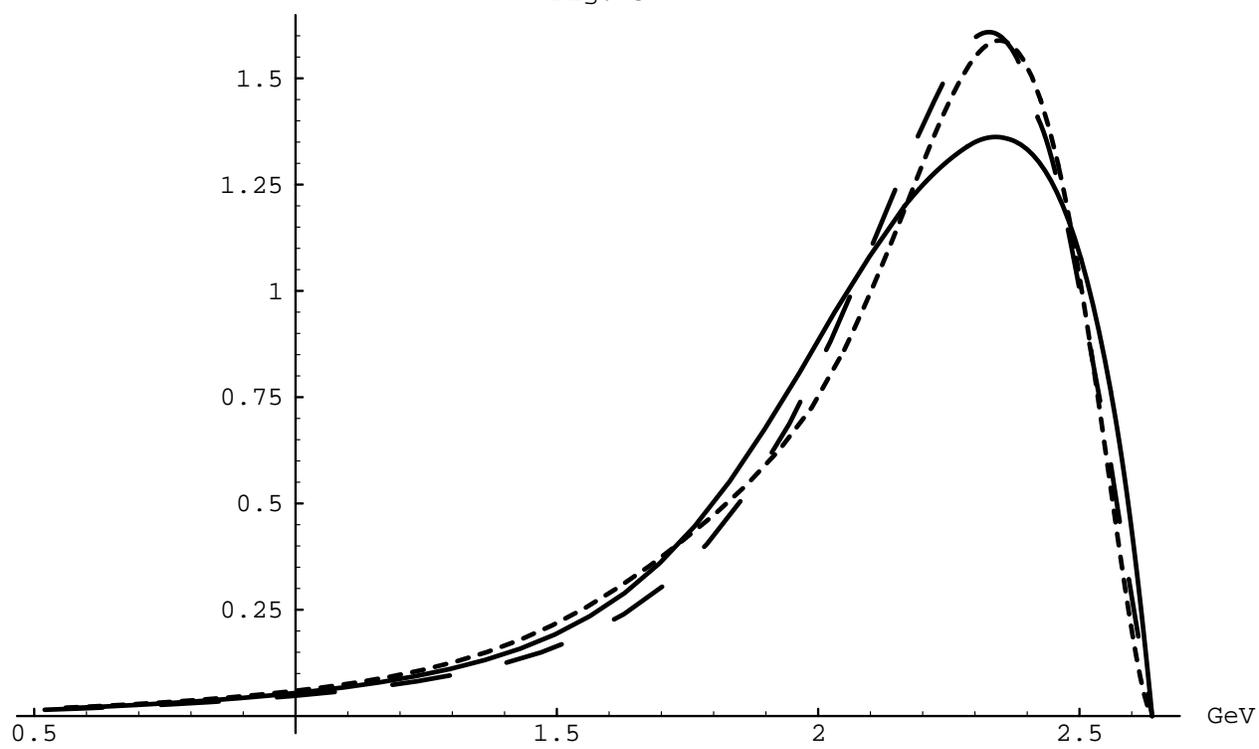
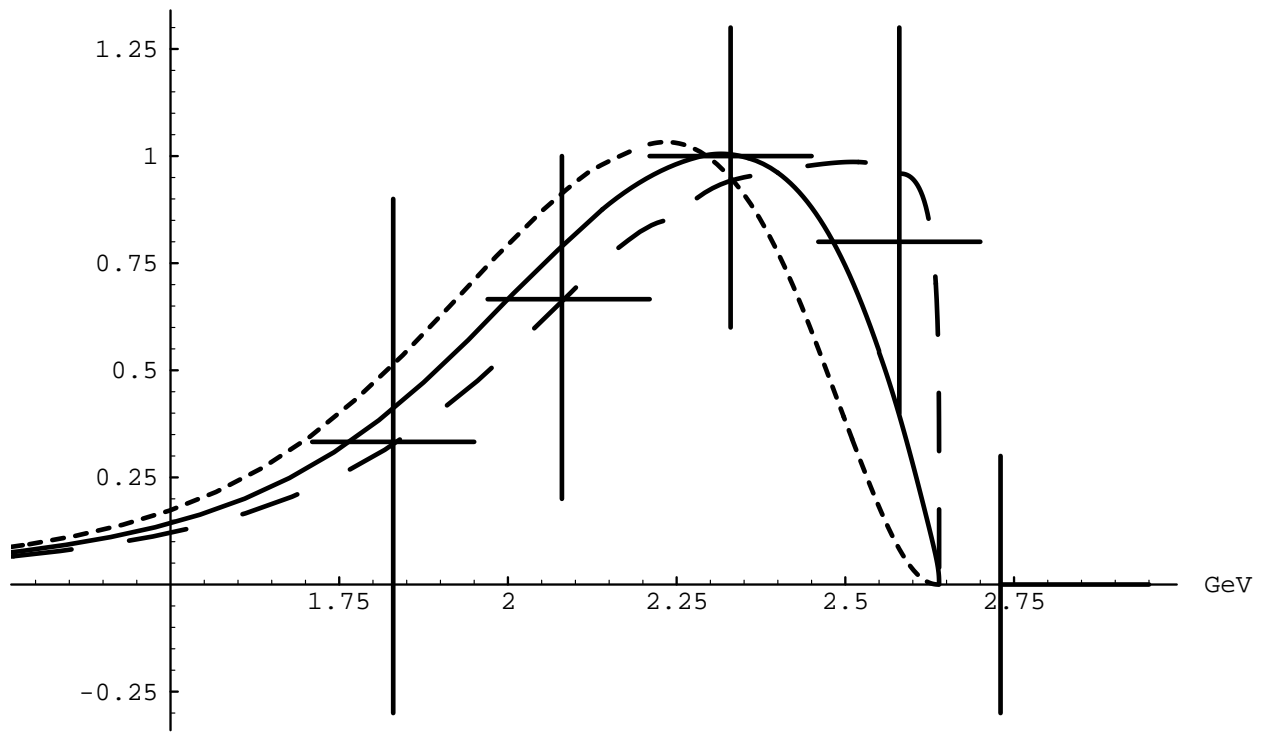
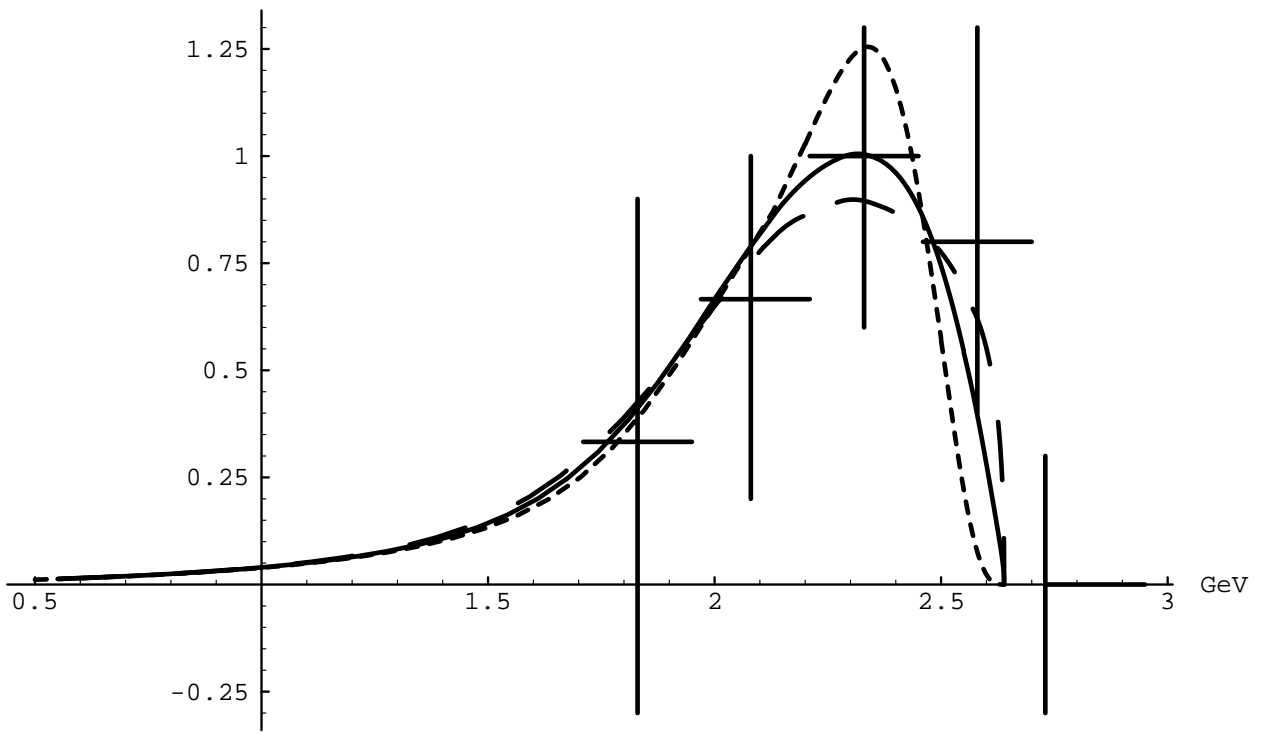
\end{document}